\documentclass[conference]{IEEEtran}
\IEEEoverridecommandlockouts
\usepackage{cite}
\usepackage{amsmath, amssymb, amsfonts}
\usepackage{algorithm}
\usepackage{algorithmic}
\usepackage{mdwlist, multicol}
\usepackage{array, booktabs, enumitem, multirow}
\usepackage{graphicx}
\usepackage{float}
\usepackage{textcomp}
\usepackage{xcolor}
\usepackage{subfigure}
\usepackage{url}
\def\BibTeX{{\rm B\kern-.05em{\sc i\kern-.025em b}\kern-.08em
    T\kern-.1667em\lower.7ex\hbox{E}\kern-.125emX}}
\begin{document}

\title{QSpeech: Low-Qubit Quantum Speech Application Toolkit}

% \author{\IEEEauthorblockN{Anonymous Authors}}

\author{\IEEEauthorblockA{Zhenhou Hong, 
Jianzong Wang\IEEEauthorrefmark{1},  
Xiaoyang Qu,
Chendong Zhao,
Wei Tao and Jing Xiao}
\IEEEauthorblockA{\textit{Ping An Technology (Shenzhen) Co., Ltd., Shenzhen, China}\\
Emails: \{hongzhenhou168,wangjianzong347,quxiaoyang343,zhaochendong343,taowei085,xiaojing661\}@pingan.com.cn}
\IEEEauthorblockN{\thanks{\IEEEauthorrefmark{1}Corresponding author: Jianzong Wang, jzwang@188.com}}
}

\maketitle

\begin{abstract}
Quantum devices with low qubits are common in the Noisy Intermediate-Scale Quantum (NISQ) era. 
However, Quantum Neural Network (QNN) running on low-qubit quantum devices would be difficult since it is based on Variational Quantum Circuit (VQC), which requires many qubits. 
% However, in the speech application, Quantum Neural Network (QNN) running on low-qubit quantum devices would be difficult since it consists of Variational Quantum Circuit (VQC), which requires many qubits. 
Therefore, it is critical to make QNN with VQC run on low-qubit quantum devices. In this study, we propose a novel VQC called the low-qubit VQC. VQC requires numerous qubits based on the input dimension; however, the low-qubit VQC with linear transformation can liberate this condition. Thus, it allows the QNN to run on low-qubit quantum devices for speech applications. Furthermore, as compared to the VQC, our proposed low-qubit VQC can stabilize the training process more. Based on the low-qubit VQC, we implement QSpeech\footnote{Our  implementation of QSpeech is publicly available at \urlstyle{tt}\url{https://github.com/zhenhouhong/QSpeech}}, a library for quick prototyping of hybrid quantum-classical neural networks in the speech field. It has numerous quantum neural layers and QNN models for speech applications. Experiments on Speech Command Recognition and Text-to-Speech show that our proposed low-qubit VQC outperforms VQC and is more stable. 
% Our implementation of QSpeech is publicly available at \urlstyle{tt}\url{https://github.com/zhenhouhong/QSpeech}.
\end{abstract}

\begin{IEEEkeywords}
Quantum Neural Networks Library, Hybrid Quantum-classical Neural Networks, Low-qubit, Variational Quantum Circuit, Speech Application
\end{IEEEkeywords}

\section{Introduction}
% \vskip -1mm
\label{sec:intro}
Recent progress in the development and commercialization of quantum computation has significantly impacted the landscape of Quantum Neural Network (QNN)~\cite{wang2021quantum,birdal2021quantum}. 
These quantum algorithms designed for the Noisy Intermediate-Scale Quantum (NISQ)~\cite{preskill2018quantum} computer are programmed with a Variational Quantum Circuit (VQC)~\cite{benedetti2019parameterized} and a classical optimizer. The approach is similar to the parameter updating in many machine learning algorithms. The QNN has been developed by some researchers in various applications~\cite{bisarya2020breast,yang2021decentralizing}.

The number of qubits in the NISQ era~\cite{preskill2018quantum} is $5$ to $50$ qubits. Several quantum devices have less than 10 qubits. The application of quantum algorithms in these quantum devices is a key challenge. Mainstream research has focused on running the quantum algorithm in low-qubit ($2$ to $10$) quantum devices~\cite{nielsen2002quantum,tavernelli2020resource}. Using a scalable trapped atomic ion system~\cite{figgatt2017complete}, the Grover Search algorithm~\cite{nielsen2002quantum} can be implemented on $3$ qubits quantum device. The Quantum Principal Component Analysis (QPCA)~\cite{lloyd2014quantum} is developed to the Resonant QPCA~\cite{li2021resonant}, which can be conducted in a low-qubit quantum device using a resonant analysis algorithm.

However, it can be challenging when QNN or hybrid quantum-classical neural networks are implemented in speech applications~\cite{qu2020evolutionary,zhang2021cacnet}. First, these algorithms with VQC in speech applications need many qubits, making them difficult to implement. Second, with numerous qubits, QNN will have the Barren Plateau problem~\cite{mcclean2018barren,zhao2021analyzing} in the training process. When restricting the number of qubits, it is challenging to assure model performance.

\begin{figure}[htbp]
\vskip -1mm
\begin{minipage}[b]{0.49\linewidth}
  \centering
  \includegraphics[width=0.95\linewidth]{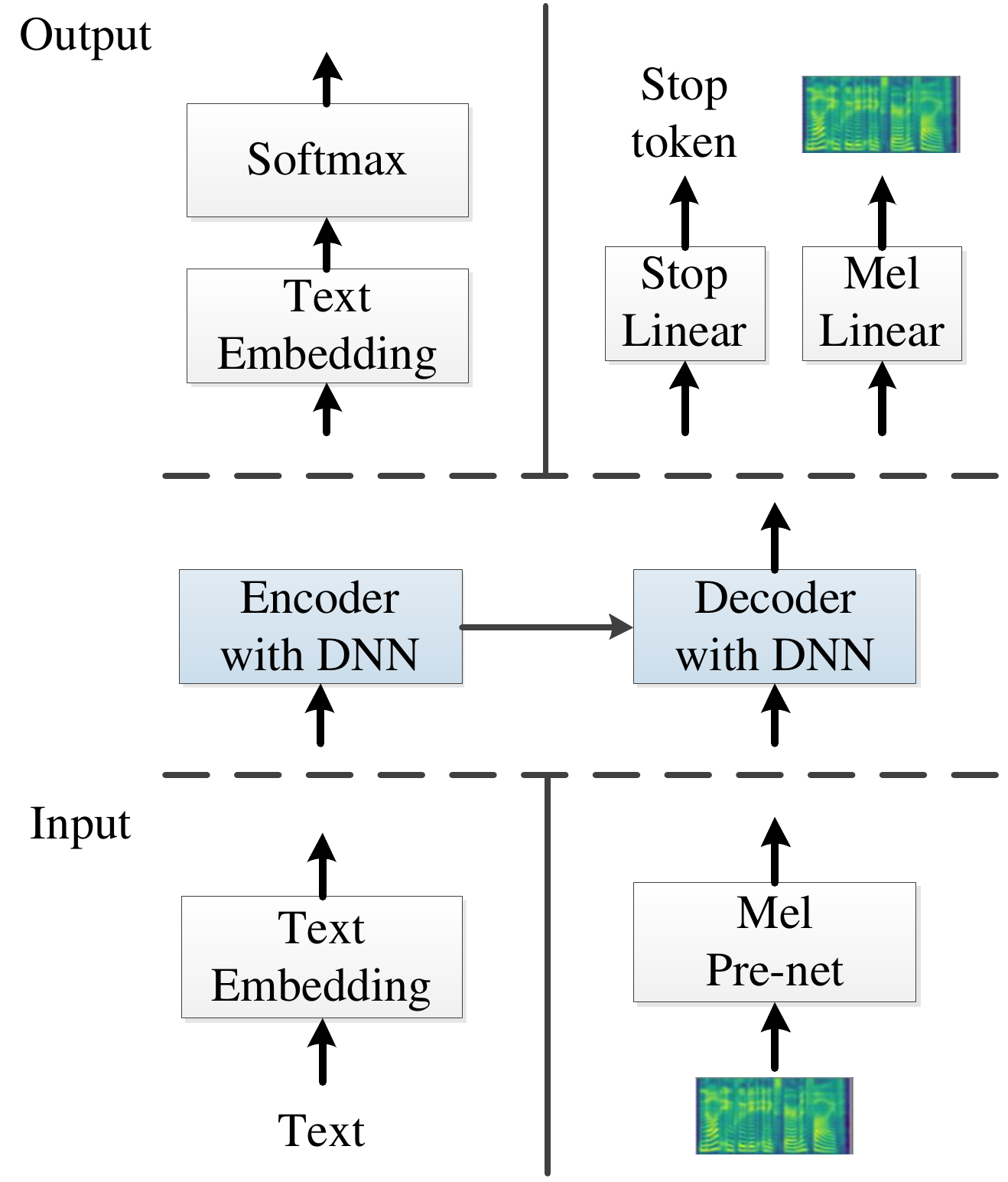}
  \\(a) The framework of speech application with DNN.
\end{minipage}
\begin{minipage}[b]{0.49\linewidth}
  \centering
  \includegraphics[width=0.95\linewidth]{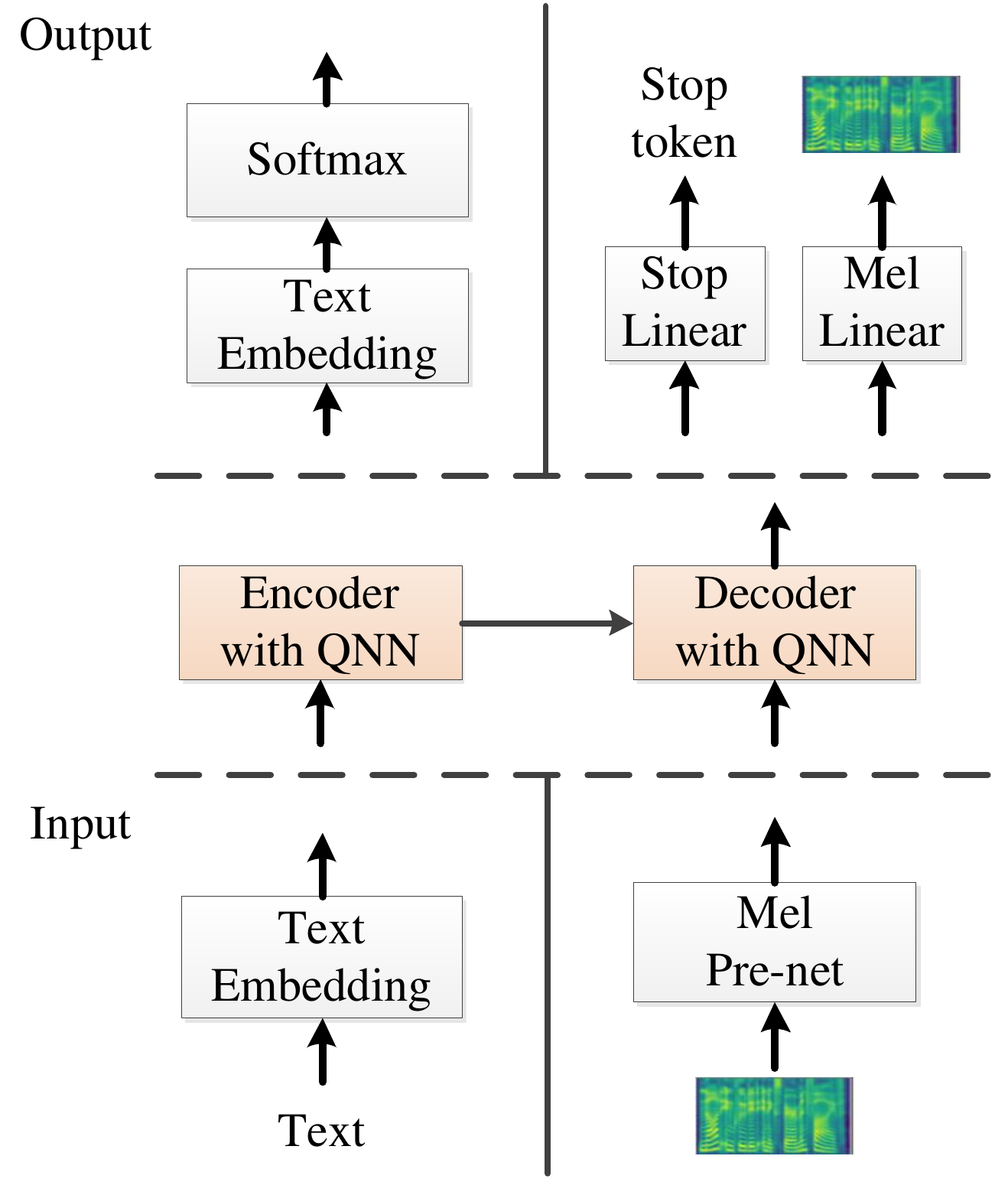}
  \\(b) The framework of speech application with QNN.
\end{minipage}
\vskip -1mm
\caption{The framework of speech application using DNN compared with QNN. The blue boxes in (a) represents the DNN modules, the orange boxes in (b) represents the QNN modules. The only difference is that DNN is replaced with QNN, with the same preprocessing and postprocessing.}
\vskip -1mm
\label{fig:speech-qspeech-process}
\end{figure}

In this study, we propose a novel VQC called low-qubit VQC that allows the QNN or hybrid quantum-classical neural networks to run in quantum devices with fewer qubits. Our proposed low-qubit VQC can be used in the neural network when high-dimensional units, such as 128 and 256, are required in the low-qubit quantum devices. Furthermore, to stabilize the convergence, we add an output clip operation in the first linear transformation in low-qubit VQC. Adding output clip operation can avoid gradient vanishing problems. Compared to VQC, the low-qubit VQC can enhance the convergence in training. Additionally, we propose a framework called QSpeech, which uses the QNN module for easy use in speech applications~\cite{hong21_interspeech,zhao2022r}. QSpeech based on the PennyLane~\cite{bergholm2018pennylane} and PyTorch~\cite{paszke2019pytorch} as backend, can define a \textit{quantum circuit} in the function with a simple Python-based user interface. We implement the neural layer functions in QSpeech such as Quantum Convolution (QConv), quantum long short term memory~\cite{chen2021quantum}, Quantum Gate Recurrent Unit (QGRU), and Quantum Attention (QAttention). Thus, the QSpeech can easily implement the hybrid quantum-classical neural network.

% \footnote{Our  implementation of QSpeech is publicly available at \urlstyle{tt}\url{https://github.com/zhenhouhong/QSpeech}}
% \footnote{The source code is publicly available at \urlstyle{tt}\url{https://github.com/zhenhouhong/QSpeech}}

Our main contributions are as follows:
\begin{itemize}[itemsep=0pt, topsep=0pt, partopsep=0pt, parsep=0pt]
\item Low-qubit VQC has been proposed, allowing QNN to be used in a low-qubit quantum device. It can stabilize the training process compared to VQC.
\item Based on low-qubit VQC, we design a library for the rapid prototyping of hybrid quantum-classical neural networks in speech applications, named QSpeech.
\item In the experiments, hybrid quantum-classical neural networks were successfully applied in speech applications.
\end{itemize}

% The implementation of QSpeech is publicly available at \urlstyle{tt}\url{https://github.com/zhenhouhong/QSpeech}.

\section{Related Works}
\subsection{Quantum Convolutional Neural Networks}
Cong et al.~\cite{cong2019quantum} propose a quantum CNN based on a quantum circuit model that incorporates the ideas of convolutional and pooling layers from classical CNNs. The proposed architecture is similarly layered; however, it applies 1D convolutions to the input quantum state (contrary to 2D/3D convolutions on images). On the input state density $\rho-{in}$, the convolutional layer is modeled as a quasi-local unitary operation. This unitary operator, denoted by $U_{i}$, is applied to several successive sets of input qubits up to a predefined depth. The pooling layer is implemented by measuring some qubits and applying unitary rotations $V_{i}$ to the nearby qubits. The observations on the qubits determine the rotation operation. This combines the functionalities of dimensionality reduction (the output of $V_{i}$ is of lower dimension) and nonlinearity (due to the partial measurement of qubits). After the required number of convolutional blocks and pooling unitaries, the unitary $F$ implements the fully connected layer. A final measurement on the output of $F$ yields the network output.

% Cong et al.~\cite{cong2019quantum} propose a quantum CNN based on a quantum circuit model that incorporates the ideas of convolutional and pooling layers from classical CNNs. The proposed architecture is similarly layered; however, it applies 1D convolutions to the input quantum state (contrary to 2D/3D convolutions on images). 
Recently, Kerenidis et al.~\cite{kerenidis2019quantum} identified the relationship between convolutions and matrix multiplications and proposed the first quantum algorithm to compute a CNN's forward pass as a convolutional product. They also proposed a quantum backpropagation algorithm to learn network parameters through gradient descent. In an application of QCNNs, Zhang et al.~\cite{zhang2019quantum} and Melnikov et al.~\cite{melnikov2019predicting} propose special convolutional neural networks for extracting features from graphs to identify those showing the quantum advantage. Hong et al.~\cite{hong2021quantum} propose a novel structure of QCNN that outperforms other DNN algorithms in predicting protein distance and contact.

\subsection{Quantum Recurrent Neural Networks}
There have also been several interesting suggestions for developing quantum variants of RNN. Hibat-Allah et al.~\cite{hibat2020recurrent} propose a quantum variant of RNNs using variational wave functions to learn the approximate ground state of a quantum Hamiltonian. Roth~\cite{roth2020iterative} propose an iterative retraining approach using RNNs for simulating bulk quantum systems via mapping translations of lattice vectors to the RNN time index. Hopfield Networks~\cite{hopfield1982neural} were a popular early type of a recurrent NN for which quantum variants have been developed in several works~\cite{tang2019experimental}.

\subsection{Quantum Learning for Speech Processing}
Li et al.~\cite{li2002quantum} propose a QNN that combines fuzzy theoretic principles with continuous digits inference. Their results show that more than $15\%$ error reduction is obtained on a speaker-independent continuous digits recognition task. Li et al.~\cite{li2009quantum} propose a QNN used in speech enhancement. The QNN can be trained on the backpropagation algorithm. Its simulation results show that the QNN outperforms the traditional spectral subtraction in speech enhancement. Fu et al.~\cite{fu2009speech} propose an improved particle swarm optimization (IPSO), called IPSO-QNN, for training the QNN. Its experimental results in speech recognition are better than other approaches. Yang et al. show that the QCNN can use in privacy-preserving learning~\cite{yang2021decentralizing}. It uses the QCNN to encode the speech feature to quantum state for data protection.

\section{Preliminaries}
\vskip -1mm
\label{sec:preliminary}
The entire quantum variational circuit~\cite{benedetti2019parameterized} consists of three components: \textit{Data Encoding Layer}, \textit{Variational Layer}, and \textit{Quantum Measurement Layer}. Notably, the number of qubits and measurements can be changed to fit the problem of interest, and the variational layer can contain many dashed boxes to increase the number of parameters, all subject to the capacity and capability of the quantum machines used in the experiments.

\textbf{Data Encoding Layer.}
We first encode a classical input vector into a quantum state in this layer, which is required for further processing. A general $N$-qubit quantum state can be represented as the following:
\begin{equation}
\left| \phi \right\rangle = \sum_{(q_{1},q_{2},\ldots,q_{N})\in \left\{ 0, 1\right\}^{N}} c_{q_{1},\ldots,q_{N}} \left| q_{1} \right\rangle \otimes \left| q_{2} \right\rangle \otimes \left| q_{3} \right\rangle \otimes \ldots \otimes \left| q_{N} \right\rangle
\end{equation}
where $c_{q_{1},\ldots,q_{N}} \in \mathbb{C}$ is the amplitude of each quantum state and $q_{i} \in \left\{ 0,1 \right\}$.
The square of the amplitude $c_{q_{1},\ldots,q_{N}}$ is the probability of measurement with the post-measurement state in $\left| q_{1} \right\rangle \otimes \left| q_{2} \right\rangle \otimes \left| q_{3} \right\rangle \otimes \ldots \otimes \left| q_{N} \right\rangle$, and the total probability should sum to 1, i.e.,
\begin{equation}
\sum_{(q_{1},q_{2},\ldots,q_{N})\in \left\{ 0, 1\right\}^{N}} \left\| c_{q_{1},\ldots,q_{N}} \right\|^{2} = 1
\end{equation}
For the quantum gates, the input will first be flattened and transformed into rotation angles. In this study, $f_{e}(\cdot)$ represent the \textit{data encoder}.

\textbf{Variational Layer.}
After encoding the classical values into a quantum state, it will be subject to a series of unitary transformations. The variational layer consists of two parts, one for entanglement and the other for rotation, which include numerous single-qubit unitary rotations $U$ parameterized by $1$ to $m$ parameters $\theta_{1},...,\theta_{m}$. The parameters labeled by $\theta_{1},...,\theta_{m}$ are the ones that will be updated by the optimization procedure. In this study, $f_{u}(\theta_{1},...,\theta_{m}; \cdot)$ represent the \textit{variational circuit}.

\textbf{Quantum Measurement(Decode) Layer.} 
We performed a quantum measurement layer to obtain the transformed data from VCs. The VCs' ensemble samplings (expectation values) were considered. Quantum simulation software (for example, PennyLane or IBM Qiskit) can be used to calculate this value deterministically. While implementing on a real quantum computer, it is necessary to prepare the same system and repeatedly use the measurements to gain adequate statistics. In this study, $f_{d}(\cdot)$ represent the \textit{measurement} layer.

\section{Proposed Methods}
\subsection{Challenges of VQC in Low-qubit Quantum Device}
\vskip -1mm
The VQC consists of \textit{data encoder}, \textit{variational circuit}, and \textit{measurement}~\cite{mitarai2018quantum,benedetti2019parameterized}. It only processes the 1D array and necessitates qubits based on the size of one dimension input. It is difficult to use the VQC in a high-dimensional input $\mathbf{x} \in \mathbb{R}^{1 \times n}$ ($n$ is 128 or 256 or even larger) because it requires $n$ qubits to encode and transform the input in VQC, which exceeds the number of qubits. Furthermore, with multiple qubits, it will be easy to cause the Barren Plateaus problem~\cite{mcclean2018barren,zhao2021analyzing} during the training process. 
\subsection{Problem Statement}
\vskip -1mm
Similar to deep learning~\cite{lecun2015deep,qu2021enhancing,liu2021automatic} in speech application, using the QNN $f_{Q}(\cdot)$ to model a function can generate $\mathbf{y}$, which represents the classify of audio or generate speech from data input $\mathbf{x}$. 
\begin{equation}
\begin{aligned}
    \mathbf{y} = f_{q}(\mathbf{W}, f_{v}(\theta), \mathbf{x})
\end{aligned}
\end{equation}
\begin{equation}
\begin{aligned}
    \mathcal{L}(\mathbf{W}, f_{v}(\theta), \mathbf{x}, \mathbf{y}_{true}) = \mathcal{L}(f_{q}(\mathbf{W}, f_{v}(\theta), \mathbf{x}), \mathbf{y}_{true})
\end{aligned}
\end{equation}
In VQC $f_{v}(\theta)$, we use the RY gate $R_{y}(\cdot)$ in \textit{Data Encoding Layer}. $\mathbf{y}_{true}$ is the true label in data.

The optimization for QNN in speech application, is similar with DNN:
\begin{equation}
\begin{aligned}
&\hat{\theta} = \mathop{\arg\min}_{\theta} \mathcal{L}(f_{q}(\mathbf{W}, f_{V}(\theta), \mathbf{x}), \mathbf{y}_{true})\\
&\hat{\mathbf{W}} = \mathop{\arg\min}_{\mathbf{W}} \mathcal{L}(f_{q}(W, f_{v}(\theta), \mathbf{x}), \mathbf{y}_{true})
\end{aligned}
\end{equation}
where the $\hat{\theta}$ and $\hat{\mathbf{W}}$ are optimized parameters in $f_{q}(\mathbf{W}, f_{v}(\theta), \mathbf{x})$. Based on PennyLane setting, VQC $f_{v}(\theta)$ can perform the optimization using gradient descent approaches. In this study, we follow the~\cite{killoran2019continuous}, using the Adam optimization method~\cite{kingma2015adam} to minimize the loss $\mathcal{L}(f_{q}(\mathbf{W}, f_{v}(\theta), \mathbf{x}), and \mathbf{y}_{true})$.

\subsection{Low-qubit Variational Quantum Circuit(VQC)}
\begin{figure}[htbp]
\vskip -2mm
\centerline{\includegraphics[width=0.9\linewidth]{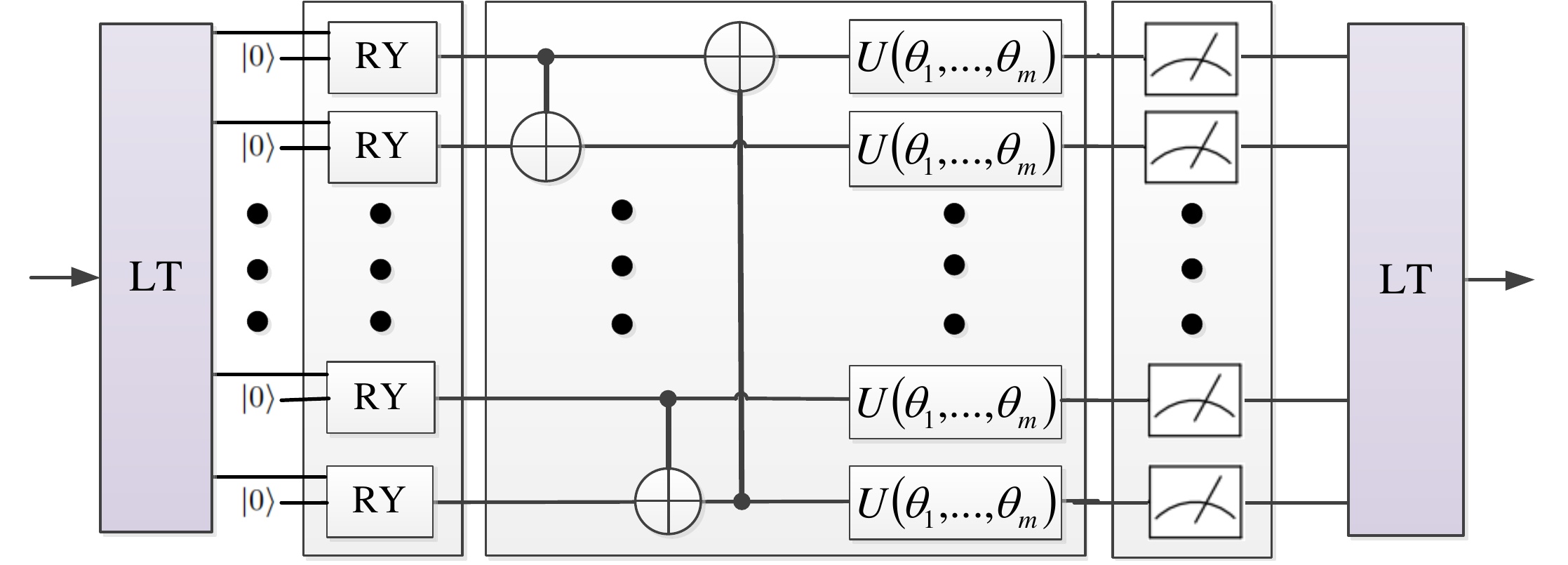}}
\vskip -2mm
\caption{The proposed low-qubit VQC.
The ``LT'' is the linear transformation, ``RY'' is the rotation gates along the y axis, $U(\theta_{1},..,\theta_{m})$ is the single-qubit unitary rotations $U$ parameterized by $1$ to $m$ parameters $\theta_{1},...,\theta_{m}$.}
\label{fig:lowqubitvqc}
\vskip -5mm
\end{figure}

% \vskip -1mm
\label{sec:lowqubit}
As shown in Fig.~\ref{fig:lowqubitvqc}, the two purple blocks are the new parts that we added in VQC. In low-qubit VQC, we add two linear transformations unlike in previous VQC~\cite{cong2019quantum,chen2021quantum},.

Consider $\mathbf{x}$ as input, let $\mathbf{x} = \{x_{1},x_{2},\dots,x_{n}\}$, $\mathbf{x} \in \mathbb{R}^{1 \times n}$. LT $f_{lt}(\mathbf{W}; \cdot)$ is linear transformation, $\mathbf{W}$ is the learnable parameter in $f_{lt}(\mathbf{W}; \cdot)$. $f_{lt}(\mathbf{W}; \cdot) : \mathbf{x} \rightarrow \mathbf{y}$. $\mathbf{q}_{1}$, $\mathbf{q}_{2}$, and $\mathbf{q}_{3}$ are representative quantum states. $\mathbf{y}_{1}$ and $\mathbf{y}_{2}$ are the outputs of first and second $f_{lt}$. $\mathbf{y}=\{y_{1},y_{2},\dots,y_{j}\}$. Then the formula of low-qubit VQC follows:
\vskip -4mm
\begin{equation}
\begin{aligned}
&\mathbf{y}_{1} = f^{1}_{lt}(\mathbf{W}; \mathbf{x})\\
&\mathbf{q}_{1} = f_{e}(\mathbf{y}_{1})\\
&\mathbf{q}_{2} = f_{u}(\theta_{1},...,\theta_{m}; c(\mathbf{q}_{1}))\\
&\mathbf{q}_{3} = f_{d}(\mathbf{q}_{2})\\
&\mathbf{y}_{2} = f^{2}_{lt}(\mathbf{W}; \mathbf{q}_{3})
\end{aligned}
\end{equation}
% \vskip -2mm
Where $\theta_{1},...,\theta_{m}$ is the parameters in $U(\theta; \cdot)$, $m$ is a positive integer. $f_{e}$, $f_{u}$, and $f_{d}$ are the \textit{data encoder}, \textit{variational circuit}, and \textit{measurement} operation. $c(\cdot)$ is the CNOT gate. $f^{1}_{lt}$ and $f^{2}_{lt}$ is the first and second linear transformation. 

The advantage of adding the linear transformation is that it can squeeze high-dimensional input in the data encoding layer or expand the output filters after measurement. Several qubits are not required since they will be squeezed with the linear transformation before the data encoding and variational layers. It can make convergence stable in the training and need fewer resources to simulate the quantum computation. Additionally, it makes low-qubit VQC process multiple channels simultaneously while extracting the features from input. It is beneficial for the QConv and QAttention layers. The QAttention layer can allow high-dimensional units such as 128 or 256 with few qubits. Thus, the Transformer with QAttention can be used in the low-qubit quantum devices for speech applications~\cite{li2021communication} such as Text-to-Speech (TTS).

\subsection{Analysis of the Low-qubit VQC}
% \vskip -1mm
We found the $f_{e}(\mathbf{y}_{1})$ contains the $\arctan(\cdot)$, its derivative may cause the gradient vanish. $f_{e}(\mathbf{y}_{1}) = R_{y}(\arctan(y_{11}))\left|0\right\rangle...R_{y}(\arctan(y_{1i}))\left|0\right\rangle$. Its derivative results of $R_{y}(\arctan(y_{1i}))$ as follow:
\begin{equation}
\begin{aligned}
&\frac{\partial R_{y}(\arctan(y_{1i}))}{\partial w_{i}} = -R_{y}(\arctan(y_{1i}))^\intercal \frac{\partial \arctan(y_{1i})}{\partial w_{i}}\\
&\frac{\partial \arctan(y_{1i})}{\partial w_{i}} = \frac{1}{1+(y_{1i})^{2}} \frac{\partial y_{1i}}{\partial w_{i}} = \frac{1}{1+(y_{1i})^{2}}x_{i}
\end{aligned}
\end{equation}
Where $y_{1i}$ is the element in $\mathbf{y}_{1}$. $y_{1i} = \mathbf{w}\mathbf{x}$. $\mathbf{w}$ is a vector in $\mathbf{W}$. 

We analysis the $x_{i} / (1 + (y_{i})^{2})$. To observe the output, we fix the $y_{i}$ and change the $x_{i}$. We can see that the $y_{i}$ cannot be too big or too small, or the gradient will vanish. Therefore, we add the output clip operation in the first linear transformation. It is possible to optimize the weight of linear transformation. The output clip's range is $[-3, 3]$.

The LT with learnable parameters can be implemented using FC. We compare the linear transformation with other subsampling approaches like bilinear interpolation. The linear transformation has two advantages: first, it can reduce the dimension and extract the features efficiently with learnable parameters; second, it can extract the features adapt to the VQC with learnable parameters. Bilinear interpolation simply reduces the dimension, causing the gradient vanish based on the derivative result. 

\subsection{Framework of QSpeech}
\label{sec:qspeech}
\begin{figure}[htbp]
% \vskip -1mm
\centerline{\includegraphics[width=0.9\linewidth]{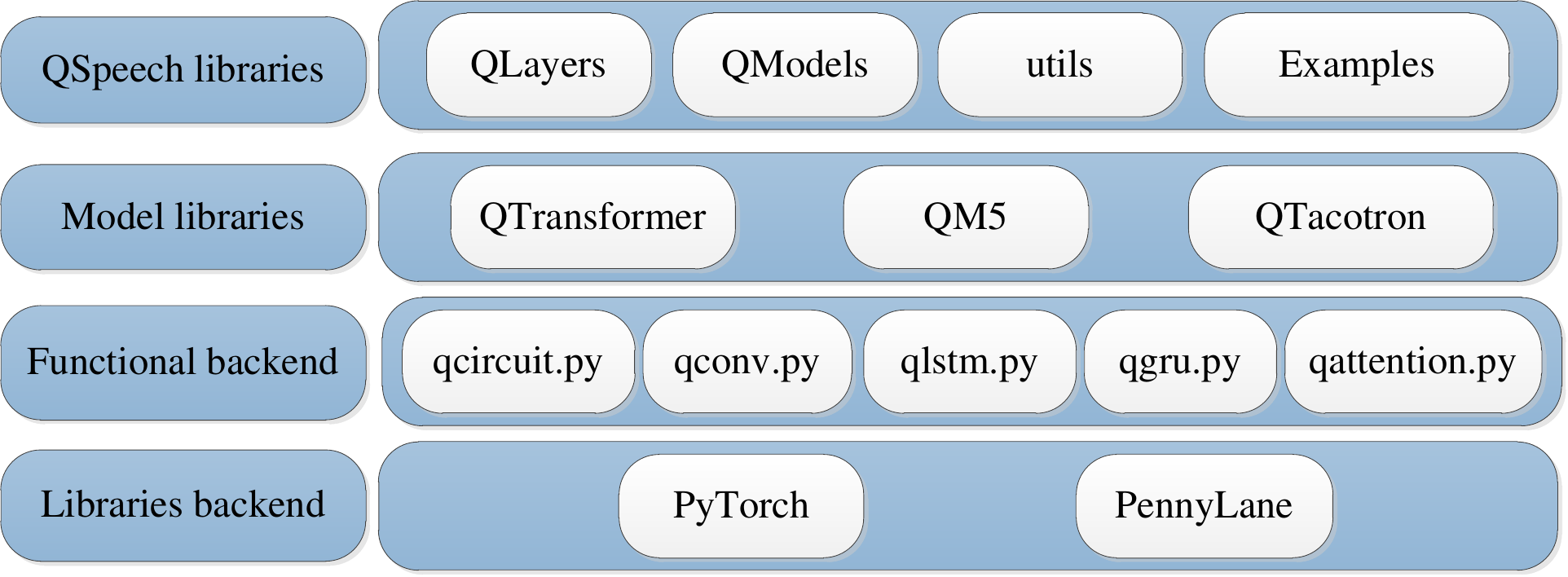}}
% \vskip -2mm
\caption{The software architecture of QSpeech. All functional backends consists of low-qubit VQC.}
% \vskip -4mm
\label{fig:architecture}
\end{figure}

% \vskip -1mm
Fig.~\ref{fig:architecture} shows the software architecture of QSpeech. In the QSpeech, the main parts of QNN training and inference are written in python, which calls PennyLane and PyTorch by switching the backend option. PennyLane allows differentiable programming of quantum computers. We also provide complete recipes for conducting speech application experiments. As shown in Fig.~\ref{fig:architecture}, the QSpeech is based on the PyTorch and PennyLane. The functional or layers backend are qcircuit, qconv, qlstm, qgru, and qattention. The functional backend builds up the model libraries: Quantum Transformer-TTS (QTransformer-TTS), Quantum M5 (QM5), and Quantum Tacotron (QTacotron). The Quantum Layers (QLayers) are the functional backend, and the Quantum Models (QModels) are the model libraries in the major libraries in QSpeech. Using the PennyLane as a Quantum Machine Learning (QML) backend can easily embed the code into Python and PyTorch.

\section{Implemented Quantum Neural Network in Speech Application}
% \vskip -1mm
\subsection{Quantum M5 For SCR}
\begin{figure}[htbp]
\vskip -3mm
\begin{minipage}[b]{\linewidth}
  \centering
  \includegraphics[width=0.9\linewidth]{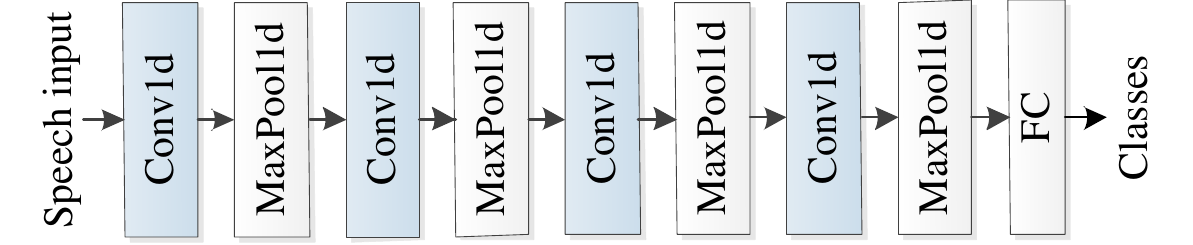}
  \\(a) M5 model.
\end{minipage}
\begin{minipage}[b]{\linewidth}
  \centering
  \includegraphics[width=0.9\linewidth]{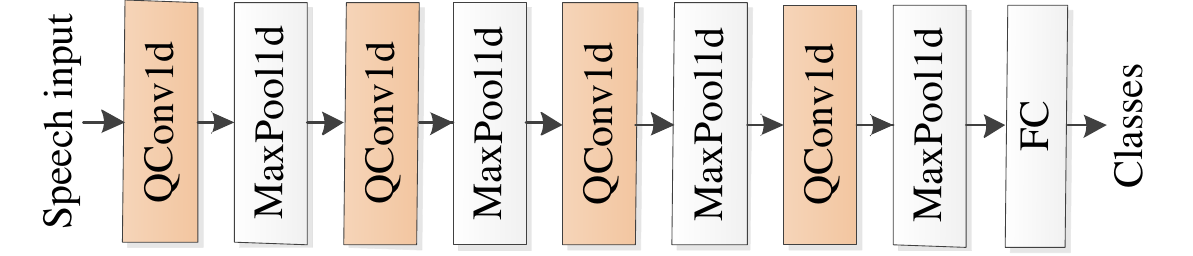}
  \\(b) QM5 model.
\end{minipage}
\vskip -2mm
\caption{The model architecture of QM5 compared with M5.}
\vskip -3mm
\label{fig:m5vsqm5}
\end{figure}

\begin{figure}
\vskip -6mm
% \centerline{\includegraphics[width=0.7\linewidth]{fig/qconv1d-05-001.pdf}}
\centerline{\includegraphics[width=7.5cm,height=2.5cm]{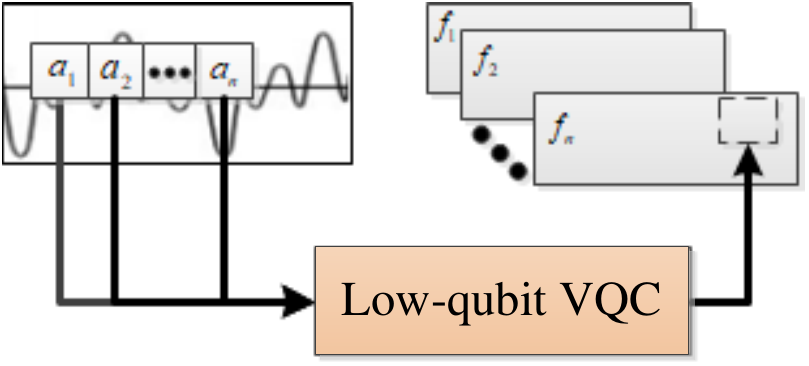}}
\vskip -2mm
\caption{The process of QConv1d operation. The $a_{1},a_{2},...,a_{n}$ means the value of input data. The $f_{1},f_{2},...,f_{n}$ means the feature maps.}
\vskip -6mm
\label{fig:qconv1d}
\end{figure}

The M5~\cite{dai2017very} model is a deep convolutional neural network (CNN) that directly uses time-domain waveforms as inputs. To extract the feature from the raw speech data, the M5 model in speech command uses one dimension convolutional layer (Conv1d). Thus, we first implement the quantum one dimension convolutional (QConv1d) layer that can use in speech commands and other speech applications~\cite{tang2021contrastive,hong2021hearing}. Then, we use the M5 model with the QConv1d layer in the speech command task, called quantum M5(QM5).

Fig.~\ref{fig:qconv1d} shows the QConv1d extract feature from raw speech data or feature map. We propose a QConv1d that uses the low-qubit VQC and help extract from a feature map, while avoiding source in the quantum simulator.

\subsection{Quantum Tacotron For TTS}
\begin{figure}[htbp]
\vskip -3mm
\begin{minipage}[b]{0.49\linewidth}
  \centering
  \includegraphics[width=0.95\linewidth]{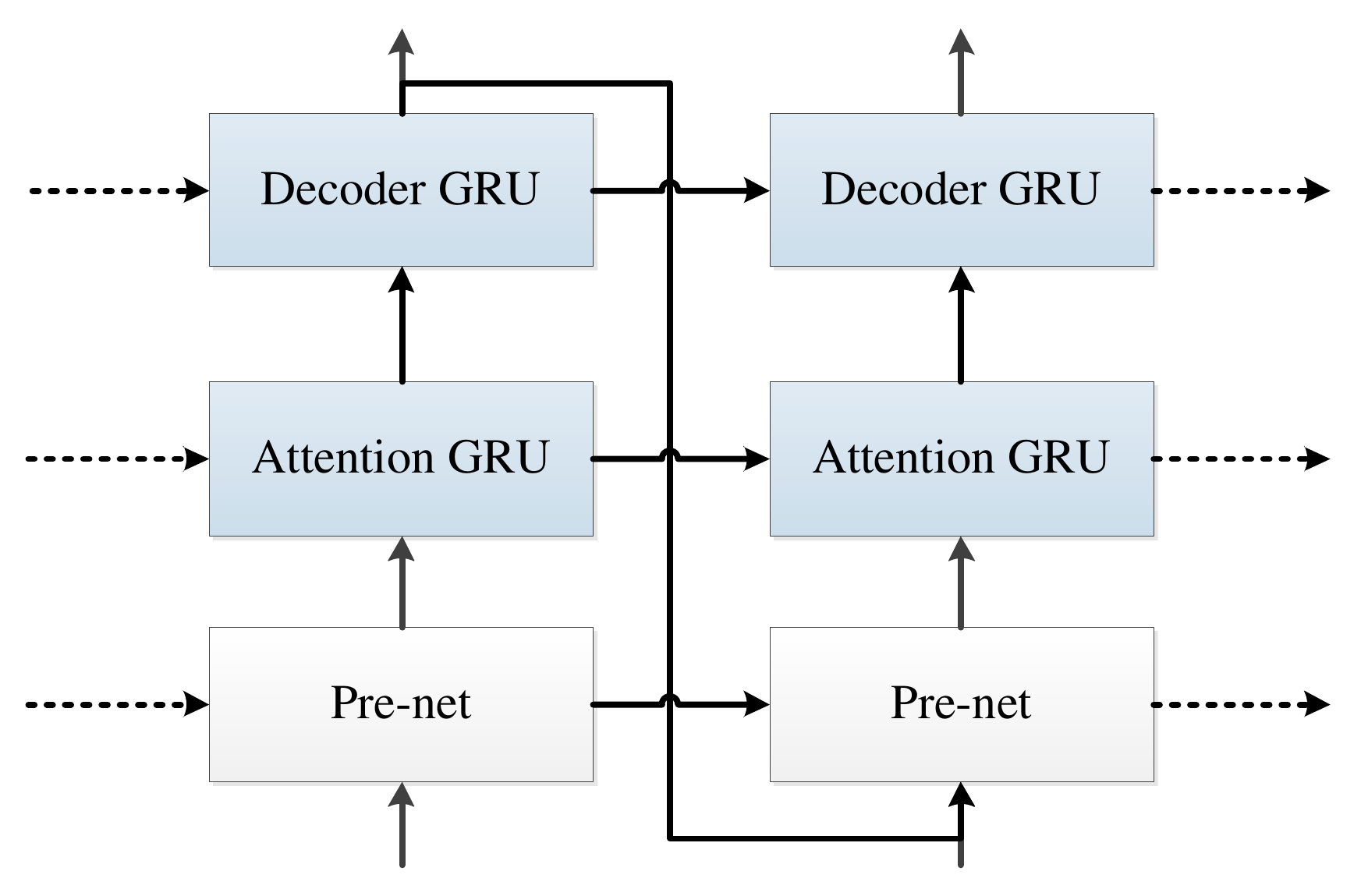}
  \\(a) Attention decoder part in Tacotron.
\end{minipage}
\begin{minipage}[b]{0.49\linewidth}
  \centering
  \includegraphics[width=0.95\linewidth]{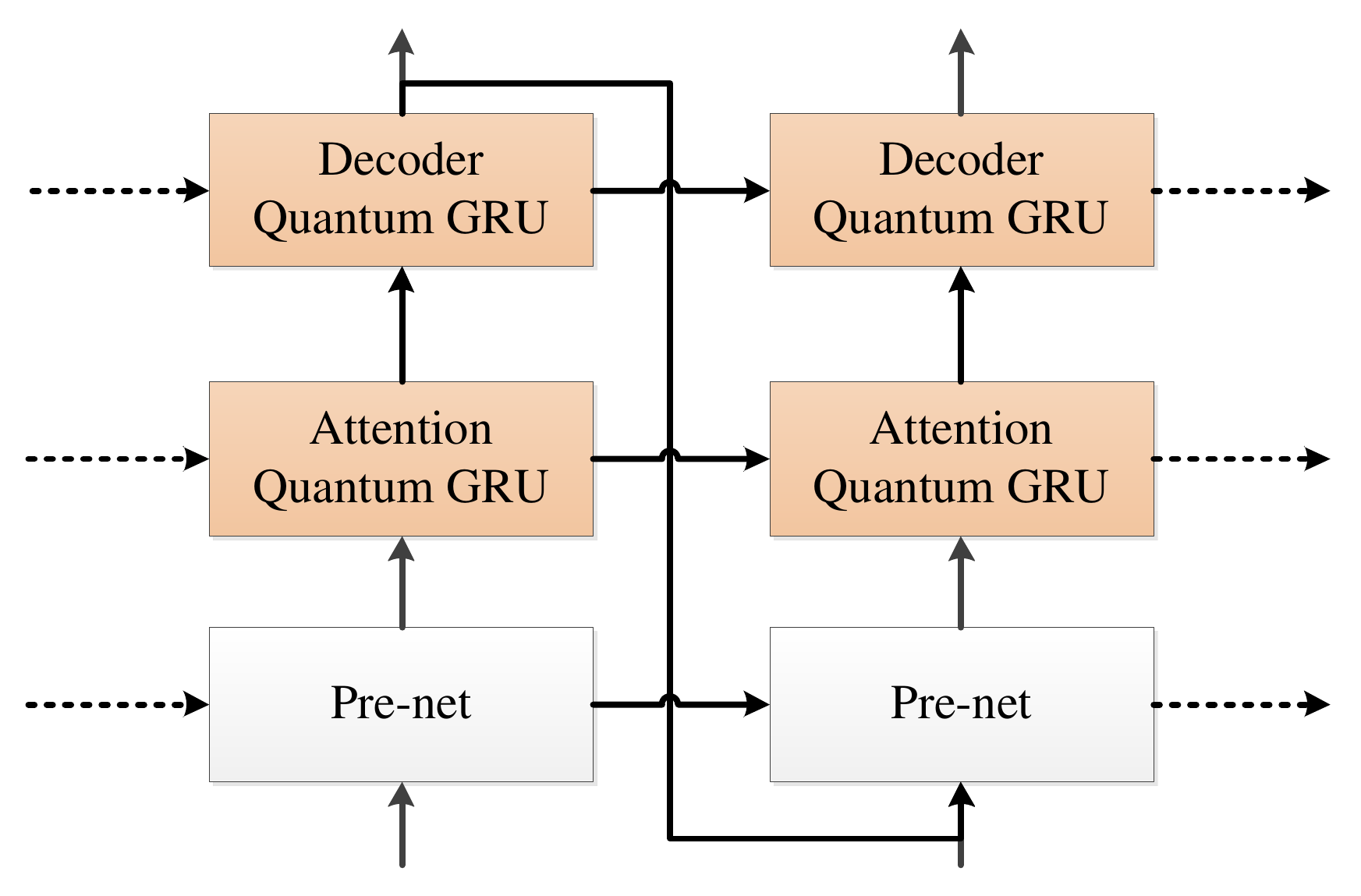}
  \\(b) Attention decoder part in QTacotron.
\end{minipage}
\vskip -2mm
\caption{The attention decoder with GRU compared with the attention decoder with quantum GRU.}
\vskip -2mm
\label{fig:qattentiondeocder}
\end{figure}

% \vskip -1mm
Tacotron~\cite{wang2017tacotron} is an end-to-end generative TTS model that synthesizes speech directly from characters. It has an Attention Decoder part, which includes a pre-net, an attention recurrent neural networks (RNN), and a decoder RNN. The RNN component uses the GRU. As shown in Fig.~\ref{fig:qattentiondeocder}, we change the original GRU to QGRU in attention GRU and decoder GRU. The others are maintained.

\begin{figure}
\vskip -4mm
% \vskip -6mm
\centerline{\includegraphics[width=0.9\linewidth]{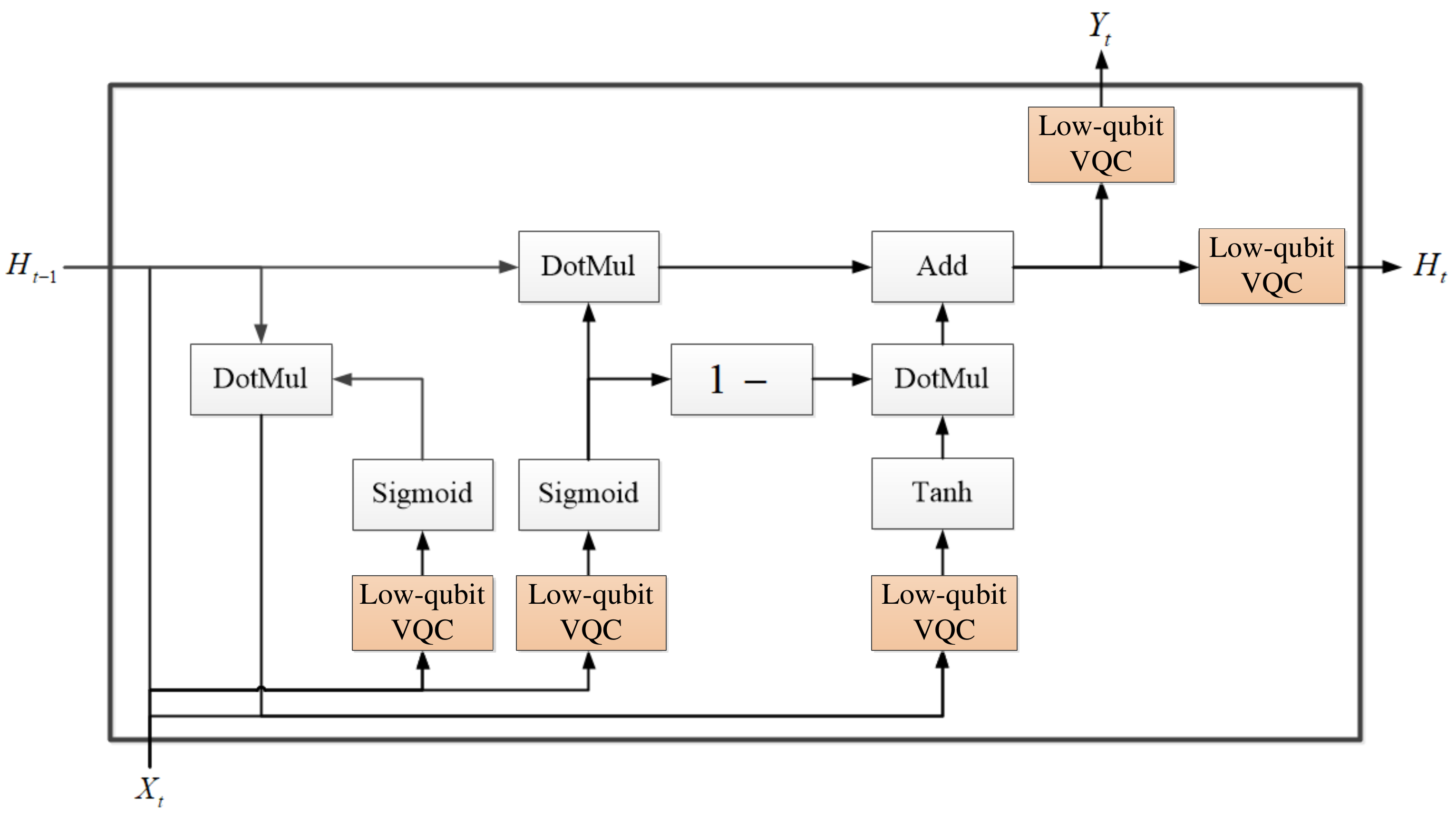}}
\vskip -2mm
% \vskip -3mm
\caption{The proposed QGRU architecture. Each low-qubit VQC as detailed in Figure~\ref{fig:lowqubitvqc}. The sigmoid and tanh blocks represent the sigmoid and the hyperbolic tangent activation function, respectively. $X_{t}$ is the input at time $t$, $H_{t}$ is for the hidden state, $Y_{t}$ is the output. ``DotMul'' represents element-wise multiplication.}
\vskip -3mm
% \vskip -4mm
\label{fig:quangru}
\end{figure}

To construct the basic unit of the proposed QGRU architecture, a QGRU cell, we stack the low-qubit VQC blocks as mentioned earlier together. In Fig.~\ref{fig:quangru}, each of the low-qubit VQC is described in the previous section (see also Fig.~\ref{fig:lowqubitvqc}). In a QGRU cell, there are five low-qubit VQC. For three low-qubit VQC in Fig.~\ref{fig:quangru} below, the input is the concatenation $v_{t}$ of the hidden state $H_{t-1}$ from the previous time step and the current input vector $X_{t}$, and the output is three vectors obtained from the measurements at the end of each low-qubit VQC. The output value can then be further processed
with two low-qubit VQC in Fig.~\ref{fig:quangru} top right, to get $Y_{t}$ and $H_{t}$.

\subsection{Quantum Transformer For TTS}

\begin{figure}[htbp]
\begin{minipage}[b]{0.49\linewidth}
  \centering
  \includegraphics[width=2.5cm,height=5cm]{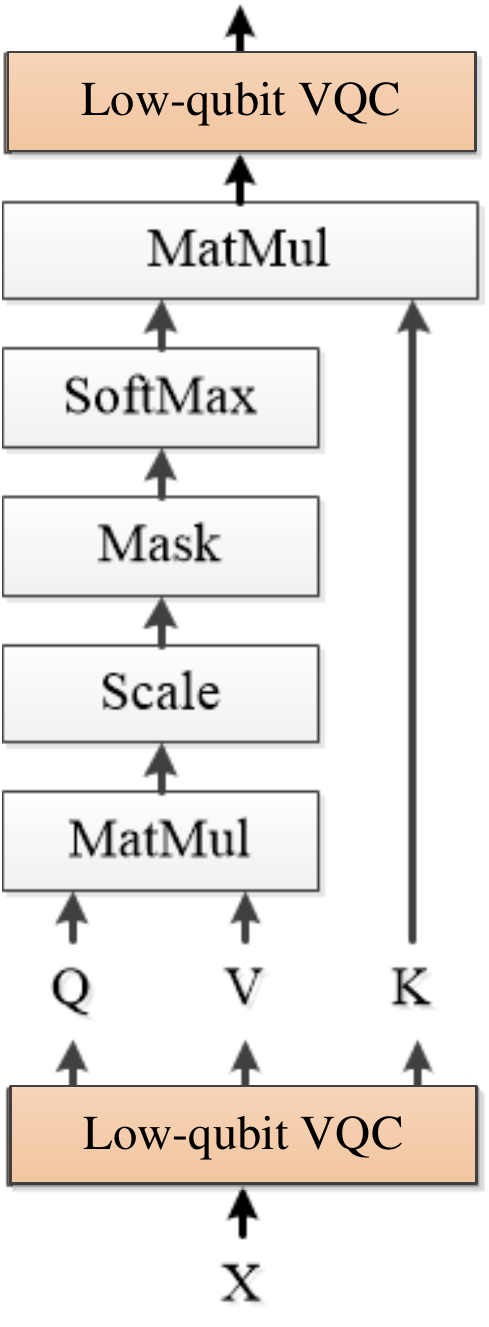}
  \\(a) Quantum self-attention with single head.
  \vskip -1mm
\end{minipage}
\begin{minipage}[b]{0.49\linewidth}
  \centering
  \includegraphics[width=3.5cm,height=5cm]{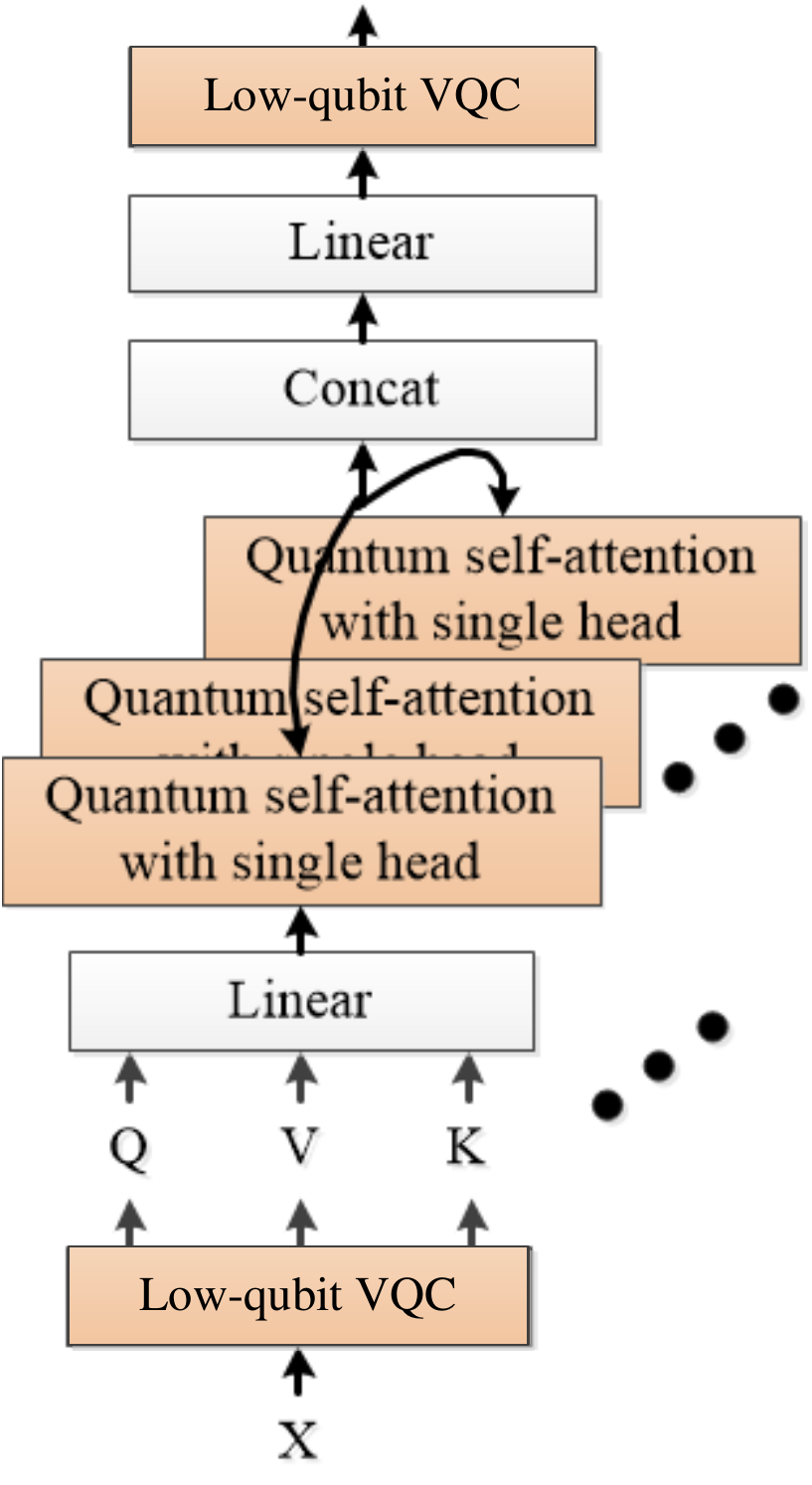}
  \\(b) Quantum self-attention with multihead.
  \vskip -1mm
\end{minipage}
% \vskip -3mm
\caption{The proposed quantum self-attention architecture with single head and multihead. The low-qubit VQC as shown in Figure~\ref{fig:lowqubitvqc}.}
\vskip -3mm
\label{fig:quanselfattention}
\end{figure}

Transformer-TTS~\cite{vaswani2017attention} is an architecture that uses two parts (Encoder and Decoder) to transform one sequence into another. Here we implement the quantum self-attention with a single head and multihead. As shown in Fig.~\ref{fig:quanselfattention}(a), adding two VQC blocks in the input and output, others are the same with Scaled-Dot Product Attention~\cite{vaswani2017attention}. The quantum self-attention with multihead as shown in Fig.~\ref{fig:quanselfattention}(b), the $Q$, $V$, $K$ computed by the VQC, input to the quantum self-attention with a single head. The final output is also computed by the VQC. The others are the same as the Multihead Attention~\cite{vaswani2017attention}.

\begin{figure}
\vskip -3mm
\centerline{\includegraphics[width=8cm,height=5cm]{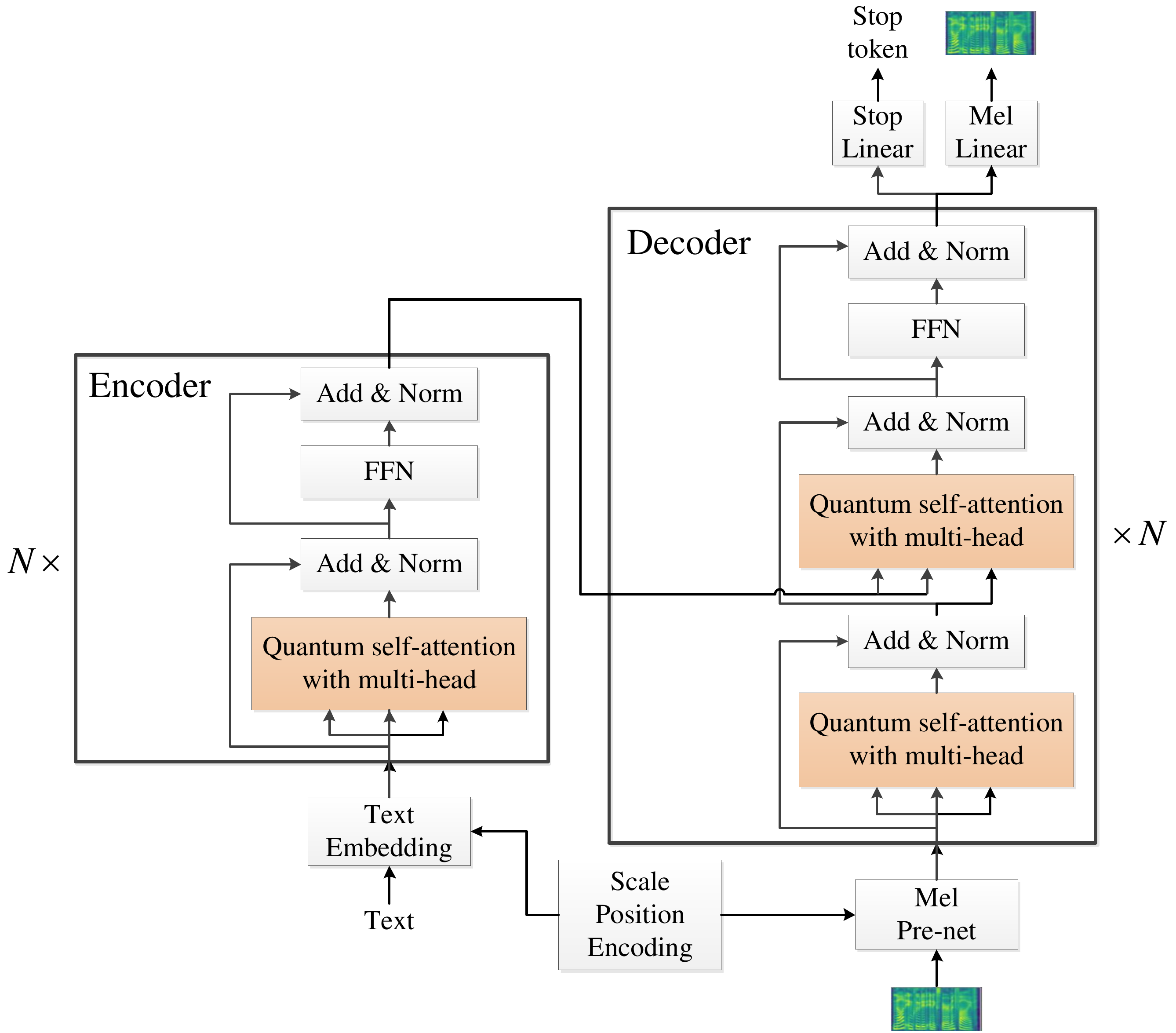}}
\vskip -2mm
\caption{The proposed QTransformer-TTS architecture.
The quantum self-attention with multihead as shown in Figure~\ref{fig:quanselfattention}(b).}
\label{fig:quantransformer}
\vskip -3mm
\end{figure}

With the quantum self-attention with multihead, the quantum Transformer-TTS (QTransformer-TTS) will also be implemented. As shown in Fig.~\ref{fig:quantransformer}, change the Multihead Attention to the quantum self-attention with multihead.

\begin{table*}[ht]
    \centering
    \caption{The Results of QM5 Compared with M5. In QM5, compared low-qubit VQC with VQC in different qubits.}
    \label{tab:resqm5}
    \begin{tabular}{c|ccccc}
       \toprule
       \textbf{Model} & \textbf{Type of VQC} & \textbf{Number of qubits} & \textbf{Batch size} & \textbf{Test accuracy} & \textbf{Training time} \\
       \hline
       \multirow{2}{*}{M5}
       & - & - & 64 & 63.75\% & - \\
       & - & - & 256 & 61.53\% & - \\
       \hline
       \multirow{12}{*}{QM5}
       & VQC & 2 & 64 & 30.57\% & 741s \\
       & VQC & 2 & 256 & 29.13\% & 2905s \\
       & Low-qubit VQC & 2 & 64 & \textbf{47.69\%} & \textbf{79s} \\
       & Low-qubit VQC & 2 & 256 & 45.61\% & 278s \\
       \cline{2-6}
       & VQC & 4 & 64 & 32.61\% & 1659s \\
       & VQC & 4 & 256 & 30.18\% & 6601s \\
       & Low-qubit VQC & 4 & 64 & \textbf{48.50\%} & \textbf{142s} \\
       & Low-qubit VQC & 4 & 256 & 47.03\% & 550s \\
       \cline{2-6}
       & VQC & 8 & 64 & 33.93\% & 3521s \\
       & VQC & 8 & 256 & 31.74\% & 13273s \\
       & Low-qubit VQC & 8 & 64 & \textbf{50.11\%} & \textbf{346s} \\
       & Low-qubit VQC & 8 & 256 & 48.97\% & 1265s \\
       \bottomrule
    \end{tabular}
    \vskip -5mm
\end{table*}

\section{Experiments}
\vskip -2mm
\label{sec:experiments}
\subsection{Dataset}
% \vskip -1mm
\textbf{LJSpeech~\cite{ito2017lj}.} 
This dataset consists of 13,100 short audio clips of a single speaker reading passages from seven nonfiction books. The texts are normalized and inserted with the beginning character (space) and end character (a period), e.g., ``There are 16 apples'' is converted to`` there are sixteen apples.''.  The LJSpeech dataset is randomly divided into two sets: 12,600 samples for training and 500 samples for testing. 

\noindent\textbf{Google Speech Command-V2~\cite{warden2018speech}.} 
For spoken word recognition, we use the 35 class setting, including the following frequent speech commands: `left', `go', `yes', `down', `up’, `on', `right', `no', `off', `stop', and so on with a total of 84,843 training examples and 11,005 testing examples.

\subsection{Settings}
% \vskip -2mm
\textbf{Quantum Environment.}
The simulation of quantum computation runs on the CPU. We limit the number of qubits to small numbers (2 to 10). The $m$ of $\theta_{1},...,\theta_{m}$ is equal to 3 in VQC and low-qubit VQC.

\noindent\textbf{Model Architecture.}
The QTransformer-TTS has two layers. It replaces self-attention with QAttention, the other settings follow~\cite{vaswani2017attention}. 
The QGRU is a substitute for GRU in QTacotron. QTacotron has same the same parameter as as~\cite{wang2017tacotron}. 
The QM5 replaces the 1D convolution (Conv1d) with 1D QConv (QConv1d).
The QTransformer-TTS, QM5, and QTacotron use the low-qubit VQC. As we discussed in Sec.~\ref{sec:lowqubit}, VQC requires similar qubits as the input dimensions, and it is considered large for quantum devices. Therefore, the low-qubit VQC in QTransformer-TTS and QTacotron, cannot be compared with VQC. The QConv1d with VQC or low-qubit VQC of QM5, in which the number of qubits is the same as the kernel size, follows the~\cite{cong2019quantum,chen2021quantum}. The linear transformation in low-qubit VQC forms a fully connected layer.

\noindent\textbf{Training.} 
Adam is the optimizer in all experiments, with $\beta_{1} = 0.9$, $\beta_{2} = 0.99$. The batch size in the training process of QTransformer-TTS and QTacotron is $2$. QM5 can be trained in a large batch size (e.g., $64$, $256$) using QConv1d with our proposed low-qubit VQC. 

\noindent\textbf{Compared Schemes.}
In the experiments, the QTransformer-TTS, QM5, and QTacotron models are compared with the original Transformer-TTS, M5, and Tacotron. The setups of training for the QTransformer-TTS, QM5, and QTacotron are the same as the original Transformer-TTS, M5, and Tacotron. It will also compare the results of different qubits in the experiments of QTransformer-TTS, QM5, and QTacotron.

\noindent\textbf{Evaluation.}
In Speech Command Recognition (SCR) experiments, the metric of accuracy is the same as~\cite{dai2017very}. We use the mean opinion score (MOS) to measure generated voices' qualities in TTS experiments. Each sentence is judged by ten males and ten females native English speakers.

\subsection{Results and Analysis}
\begin{figure}[htbp]
% \vskip -2mm
\begin{minipage}[htbp]{0.49\linewidth}
  \centering
  \includegraphics[width=1.0\linewidth]{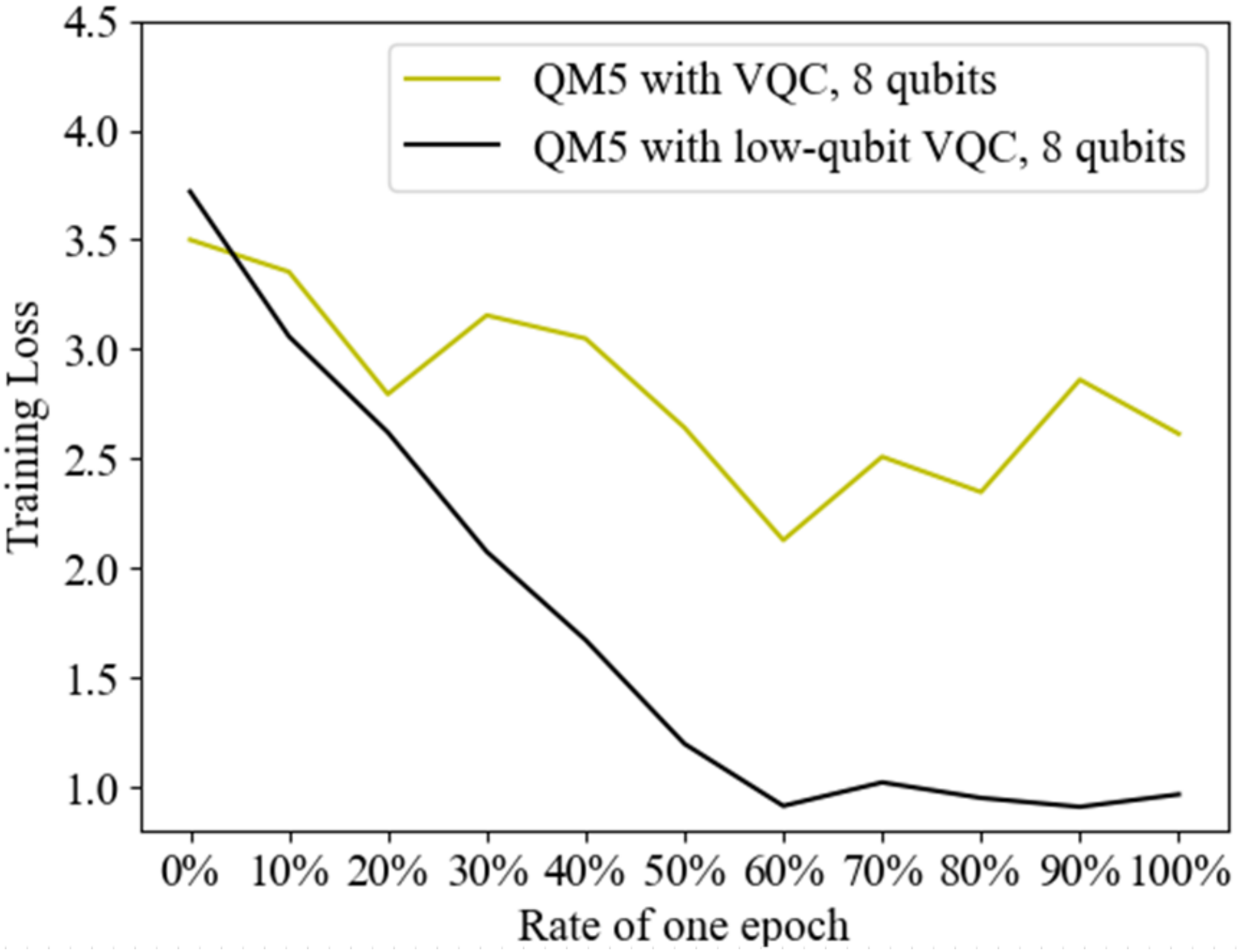}
  \\(a) Loss of VQC and low-qubit VQC in QM5.
\end{minipage}
\begin{minipage}[htbp]{0.49\linewidth}
  \centering
  \includegraphics[width=1.0\linewidth]{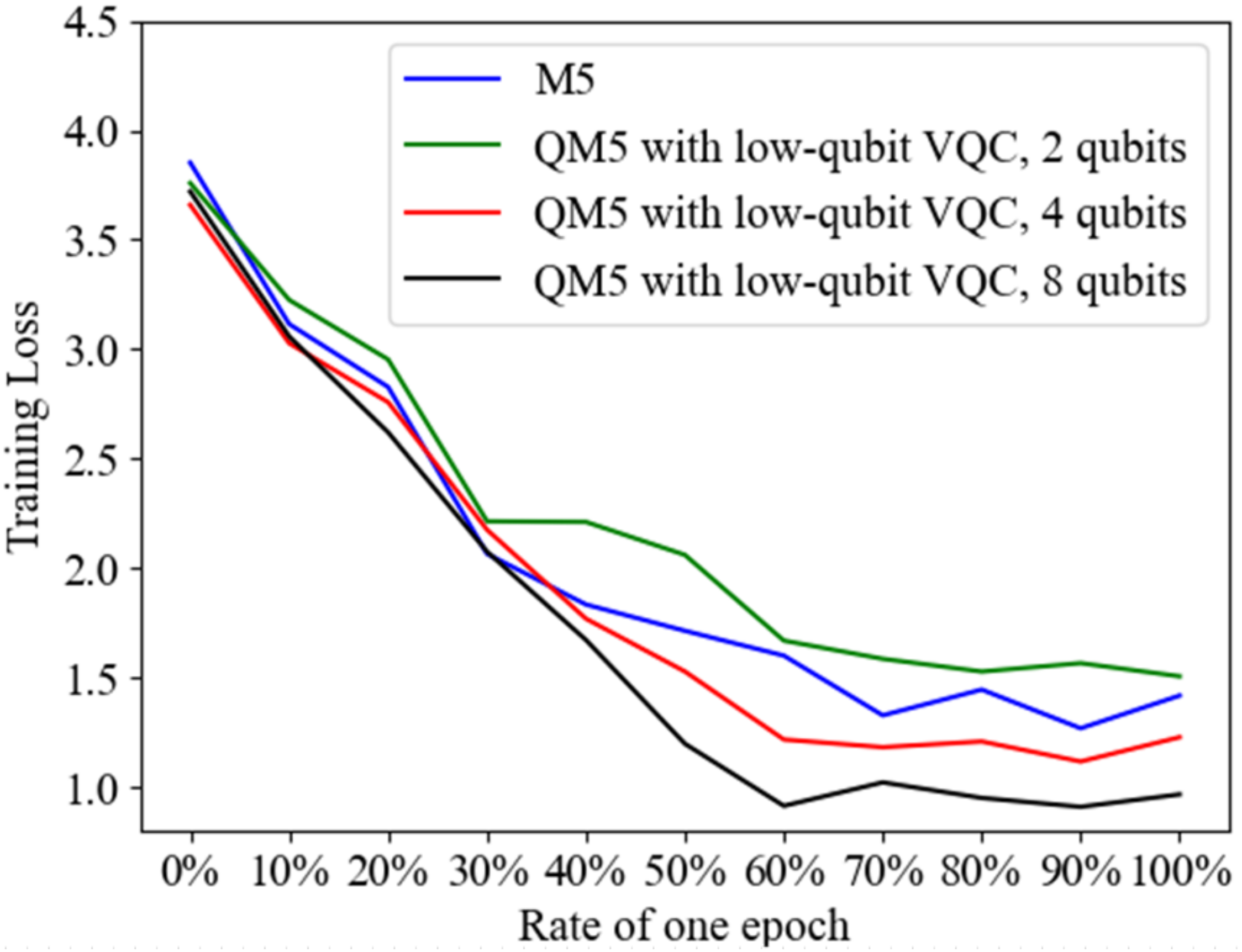}
  \\(b) Loss of M5 and QM5(with low-qubit VQC).
\end{minipage}
% \vskip -2mm
\caption{The loss curves of M5 and QM5. (a) The training losses of QM5 with VQC and low-qubit VQC. (b) The training loss curve of QM5 with different numbers of qubits, compared with M5.}
% \vskip -2mm
\label{fig:lossm5qm5}
\end{figure}

% \vskip -1.5mm
\textbf{QM5.}
In Fig.~\ref{fig:lossm5qm5}(a), the yellow curve is the training loss of QM5 with VQC. Notably, it cannot stably converge as the iterative process increases. The black curve, which is the QM5 with low-qubit VQC, can train in convergence and stability, indicating that using low-qubit VQC can handle the Barren Plateaus challenge. In Fig.~\ref{fig:lossm5qm5}(b), the training of QM5 with low-qubit VQC can improve convergence and stability compared to the M5. With more qubits in QM5, the QM5 model will converge faster.

In Table~\ref{tab:resqm5}, with additional qubits, the QM5 obtains better performance. Compared with VQC, the low-qubit VQC is $16\%$ higher. Table~\ref{tab:resqm5} shows that the low-qubit VQC is faster than the VQC. For the same batch size, the low-qubit VQC can reduce the training by $10$ times compared to the VQC. Table~\ref{tab:resqm5} also shows the accuracy between the M5 and the QM5. The QM5, when using the low-qubit VQC, can gain $16\%$ improvement in accuracy compared to the VQC, indicating that the QM5 can work in the SCR task.

\begin{figure}[htbp]
% \vskip -2mm
\hspace{1mm}
\begin{minipage}[b]{0.49\linewidth}
  \centering
  \includegraphics[width=0.9\linewidth]{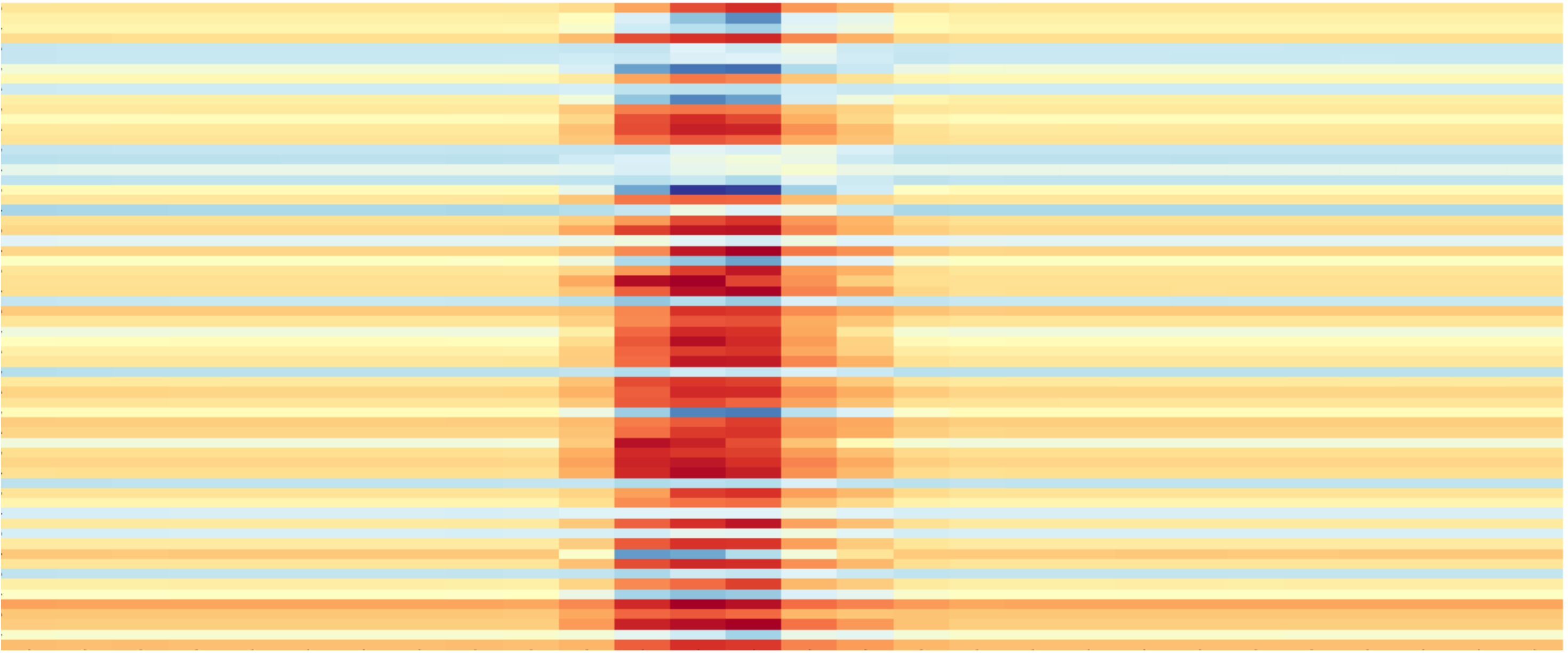}
  \\(a) Feature map of M5.
\end{minipage}
\hspace{-4mm}
\begin{minipage}[b]{0.49\linewidth}
  \centering
  \includegraphics[width=0.9\linewidth]{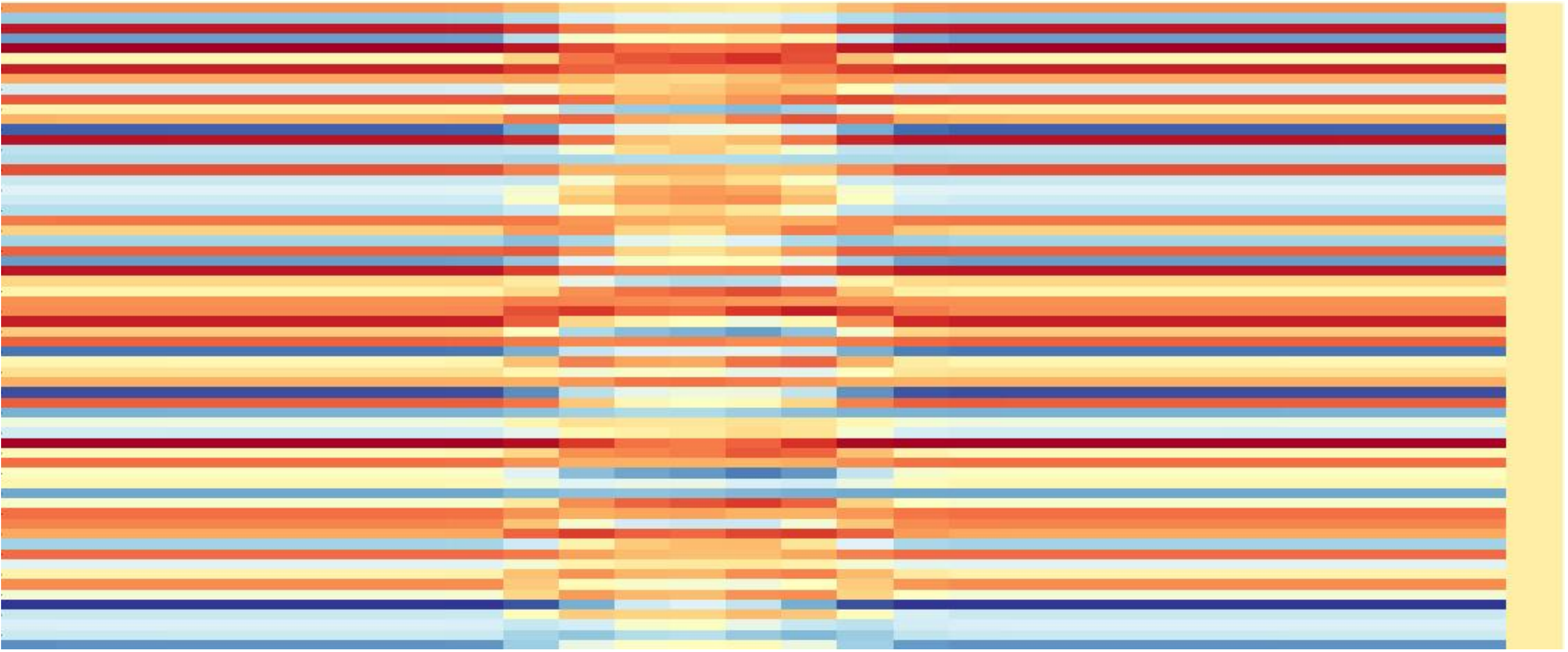}
  \\(b) Feature map of QM5.
\end{minipage}
% \vskip -2mm
\caption{The feature map output from M5 and QM5.(a) The feature map output from Conv1d in M5.(b) The feature map output from QConv1d in QM5.}
\vskip -3mm
\label{fig:fm5qm5}
\end{figure}

As shown in Fig.~\ref{fig:fm5qm5}, the QConv1d can extract the feature from an input like Conv1d. The feature pattern extracted by QConv1d is similar to that by Conv1d.

\begin{figure}[htbp]
\vskip -2mm
\begin{minipage}[b]{0.49\linewidth}
  \centering
  \includegraphics[width=1.0\linewidth]{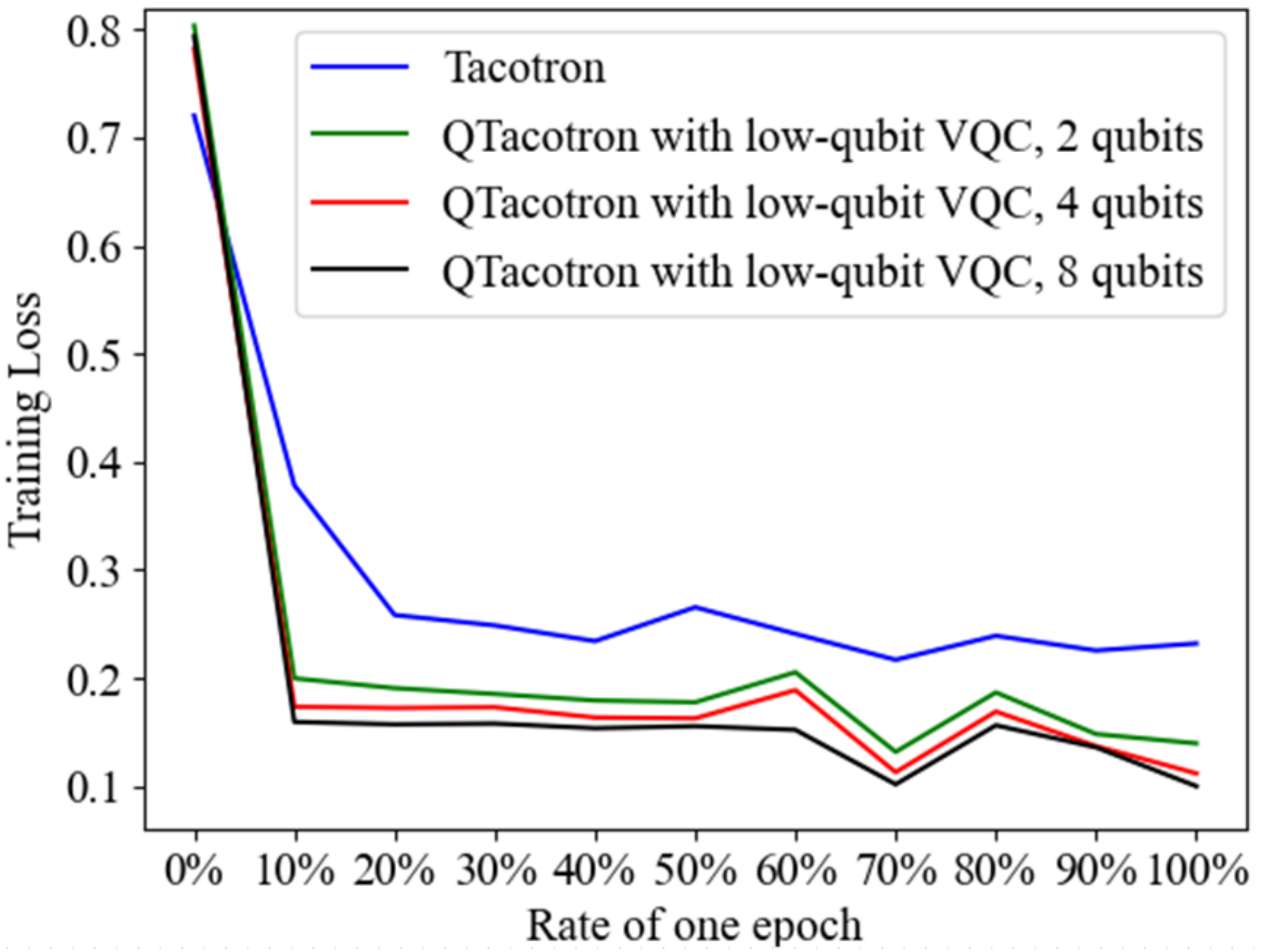}
  \\(a) Loss of Tacotron and QTacotron.
\end{minipage}
\begin{minipage}[b]{0.49\linewidth}
  \centering
  \includegraphics[width=1.0\linewidth]{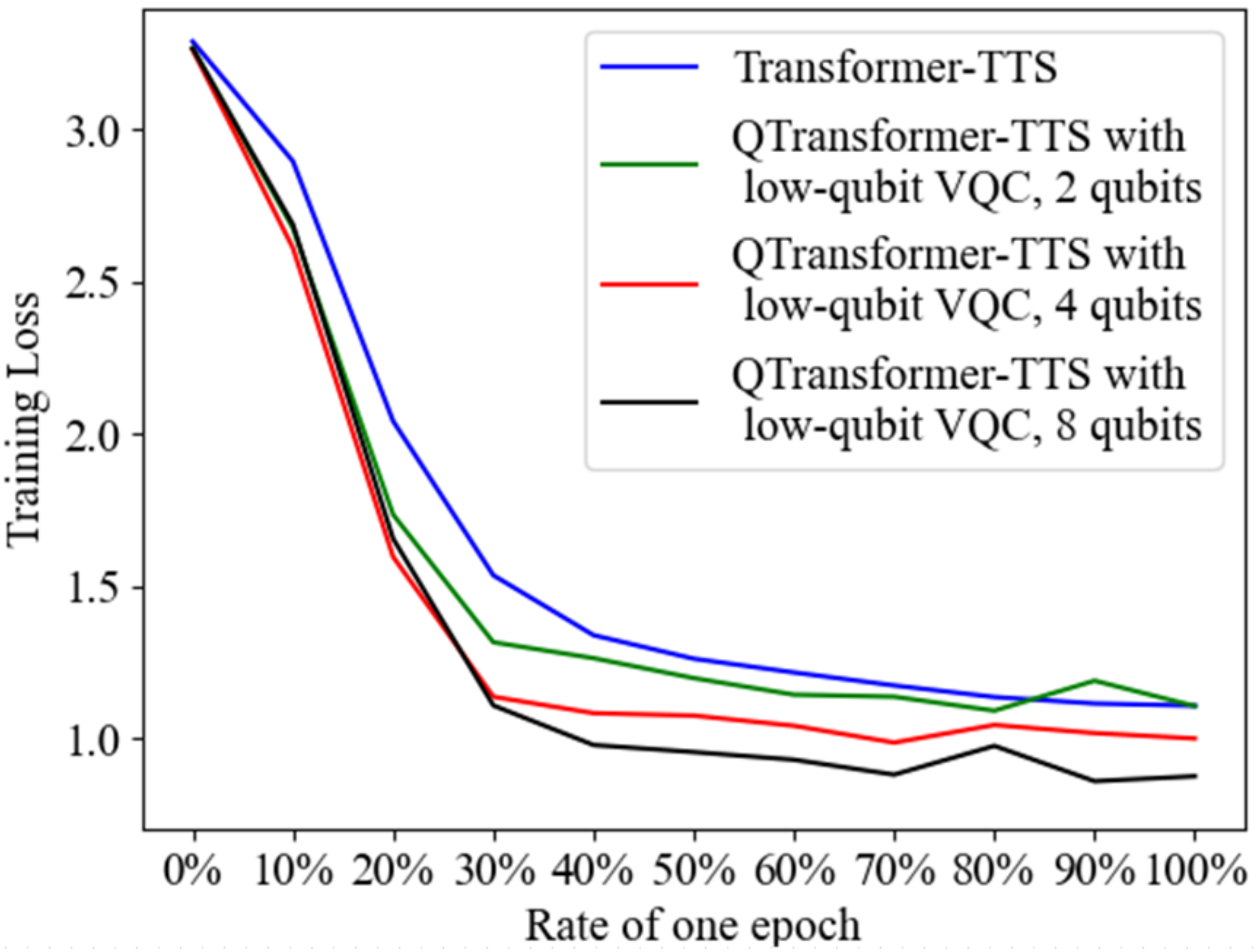}
  \\(b) Loss of Transformer-TTS and QTransformer-TTS.
\end{minipage}
\vskip -2mm
\caption{The loss curves of Tacotron, QTacotron, Transformer-TTS, and QTransformer-TTS. For QTacotron and QTransformer-TTS, both are compared with different number of qubits. (a) The loss curves of Tacotron and QTacotron. (b) The loss curves of Transformer-TTS and QTransformer-TTS.}
\vskip -2mm
\label{fig:fqtata}
\end{figure}

\noindent\textbf{QTacotron.}
As shown in Fig.~\ref{fig:fqtata}(a), the QTacotron with the low-qubit VQC converges faster in training than Tacotron (the blue curve), indicating that the low-qubit VQC can operate well in QGRU, like the low-qubit VQC in QConv1d. The $0\%$ to $20\%$ of the training process indicates that the QTacotron convergence is faster than that of the QTacotron. Following the training steps, the QTacotron has less oscillation than the Tacotron. In Fig.~\ref{fig:fqtata}(a), QTacotron with different qubits has a similar loss convergence. However, it shows that with more qubits, the QTacotron can converge better and with less loss.

\begin{table}[ht]
    \centering
    \vskip -2mm
    \caption{Results of Tacotron, QTacotron, Transformer-TTS, and QTransformer-TTS. QTacotron and QTransformer-TTS run in different numbers of qubits with low-qubit VQC.}
    % \vskip -2mm
    \label{tab:resqtaqt}
    \begin{tabular}{c|ccc}
      \toprule
      \textbf{Model} & \textbf{Type of VQC} & \textbf{\#Qubits} & \textbf{MOS} \\
      \hline
      Tacotron & - & - & 3.09 \\
      \hline
      \multirow{3}{*}{QTacotron}
      & Low-qubit VQC & 2 & 2.22 \\
      & Low-qubit VQC & 4 & 2.56 \\
      & Low-qubit VQC & 8 & \textbf{2.67} \\
      \hline
      Transformer-TTS & - & - & 2.91 \\
      \hline
      \multirow{3}{*}{QTransformer-TTS}
      & Low-qubit VQC & 2 & 2.19 \\
      & Low-qubit VQC & 4 & 2.43 \\
      & Low-qubit VQC & 8 & \textbf{2.52} \\
      \bottomrule
    \end{tabular}
    \vskip -2mm
\end{table}

Table~\ref{tab:resqtaqt} shows the MOS with Tacotron and QTacotron. QTacotron's performance will improve with more qubits of the low-qubit VQC. Therefore, the QTacotron can achieve better results when using the low-qubit VQC. With eight qubits of the low-qubit VQC, the MOS of QTacotron is about $2.67$, indicating that the QTacotron with low-qubit VQC can run in the TTS task.

\begin{figure}[htbp]
\vskip -3mm
\begin{minipage}[b]{1.0\linewidth}
  \centering
  \includegraphics[width=0.9\linewidth]{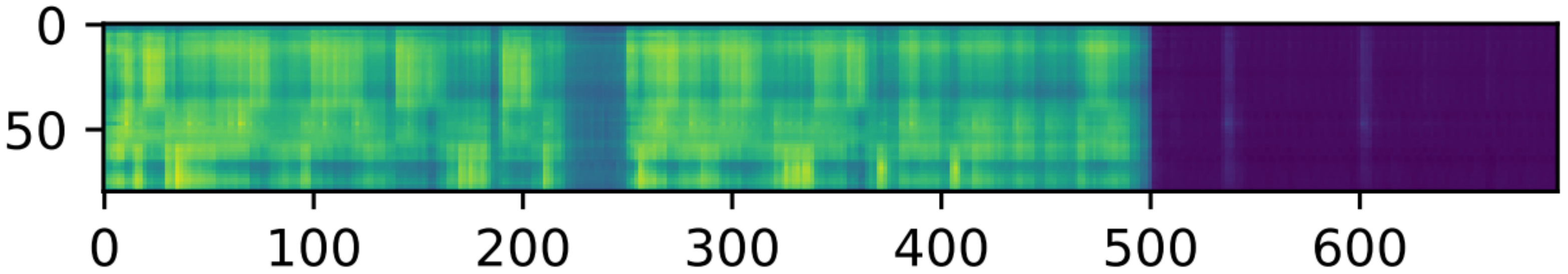}
  \\(a) Mel-spectrogram of Tacotron.
\end{minipage}
\begin{minipage}[b]{1.0\linewidth}
  \centering
  \includegraphics[width=0.9\linewidth]{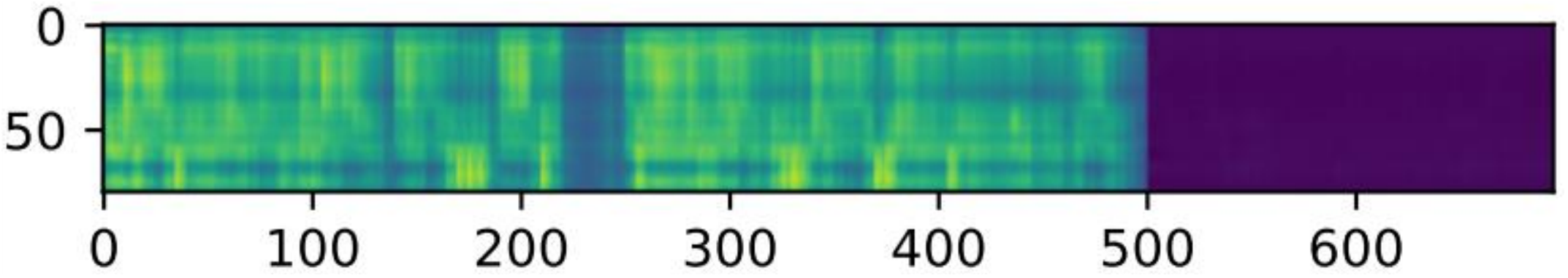}
  \\(b) Mel-spectrogram of QTacotron.
\end{minipage}
\begin{minipage}[b]{1.0\linewidth}
  \centering
  \includegraphics[width=0.9\linewidth]{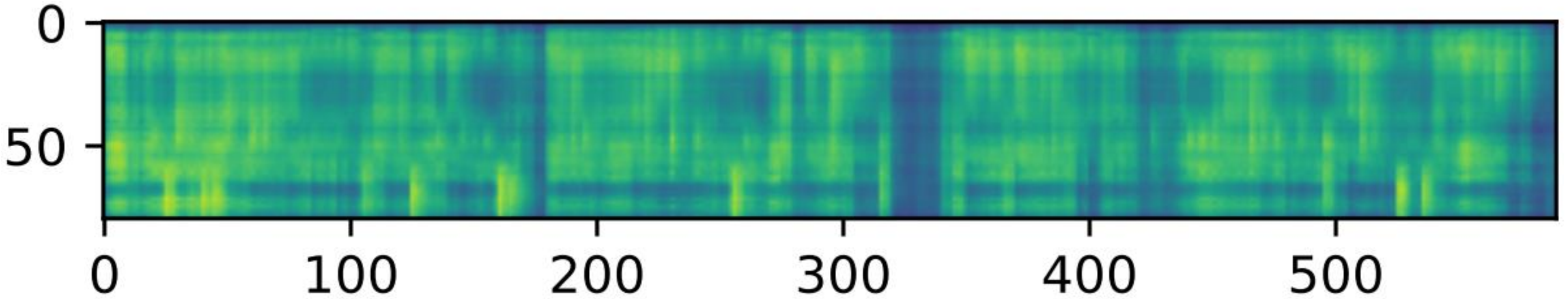}
  \\(c) Mel-spectrogram of Transformer-TTS.
\end{minipage}
\begin{minipage}[b]{1.0\linewidth}
  \centering
  \includegraphics[width=0.9\linewidth]{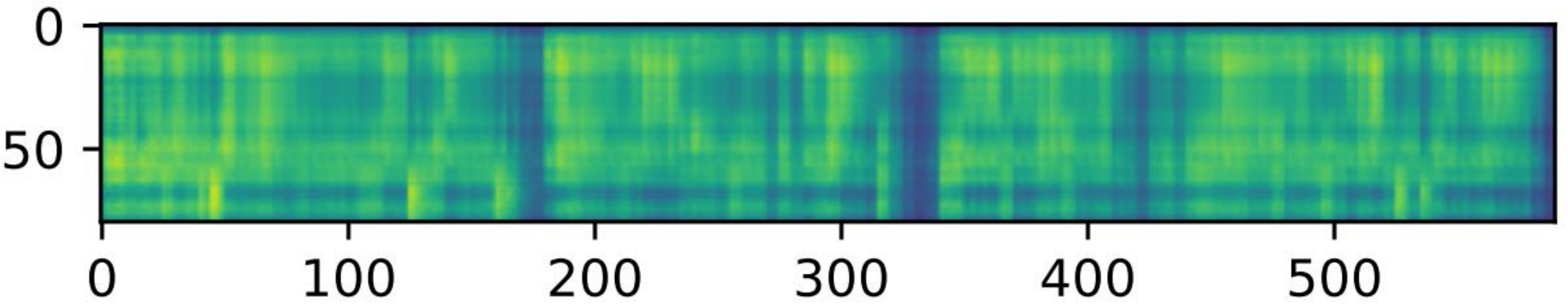}
  \\(d) Mel-spectrogram of QTransformer-TTS.
\end{minipage}
\vskip -2mm
\caption{The mel-spectrogram between the Tacotron, QTacotron, Transformer-TTS, and QTransformer-TTS. (a) Mel-spectrogram generate by Tacotron. (b) Mel-spectrogram generate by QTacotron. (c) Mel-spectrogram generate by Transformer-TTS. (d) Mel-spectrogram generate by QTransformer-TTS.}
\vskip -3mm
\label{fig:fqtt}
\end{figure}

In addition to the training process and performance, we provide the mel-spectrogram obtained from the Tacotron and QTacotron. As shown in Fig.~\ref{fig:fqtt}(a)(b), similar mel-spectrograms are obtained from the Tacotron and QTacotron. We can observe that the QTacotron with low-qubit  can achieve the approximate performance as Tacotron.

\noindent\textbf{QTransformer-TTS.}
In the training process, QTransformer-TTS outperforms Transformer-TTS, as shown in Fig.~\ref{fig:fqtata}(b). Compared to the training loss curve in Fig.~\ref{fig:fqtata}(b), the QTransformer-TTS converges faster than the Transformer-TTS between $0\%$ and $40\%$ of the training process. Furthermore, QTransformer-TTS, like the QTacotron, gives similar loss convergence in different qubits. With more qubits of the low-qubit VQC, the QTransformer-TTS obtains lower loss in the training process, indicating that the low-qubit VQC can work well in QAttention, like the low-qubit VQC in QConv1d and QGRU.

Table~\ref{tab:resqtaqt} also shows the MOS with Transformer-TTS and QTransformer-TTS. We can see that QTransformer-TTS performs better when using the low-qubit VQC. With more qubits of the low-qubit VQC, QTransformer-TTS will get better performance. With eight qubits of the low-qubit VQC, the MOS of QTransformer-TTS is about $2.52$. It means that the QTransformer-TTS with low-qubit VQC can run in the TTS task with a low-qubit quantum device.

Furthermore, in Fig.~\ref{fig:fqtt}(c)(d), we show the mel-spectrogram generated by the Transformer-TTS and QTransformer-TTS. The mel-spectrogram of the Transformer-TTS and QTransformer-TTS is similar. We conclude that QTransformer-TTS with low-qubit VQC can achieve the approximate performance as Transformer-TTS.

\noindent\textbf{Ablation Study.}
The low-qubit VQC is compared in two scenarios: first, the LT is implemented by FC layer or bilinear interpolation; second, the LT layer with or without output clip operation. Table~\ref{tab:ablation} shows the results.

\begin{table}[ht]
    \centering
    \vskip -2mm
    \caption{Compared results of QM5, QTacotron, and QTransformer-TTS. QTacotron and QTransformer-TTS run with low-qubit VQC in different settings.}
    \vskip -2mm
    \label{tab:ablation}
    \begin{tabular}{c|ccc}
      \toprule
      \textbf{Model} & \textbf{Methods of LT} & \textbf{Results} \\
      \bottomrule
      \multirow{2}{*}{QM5}
      & use bilinear interpolation & 37.51\% \\
      & use FC & \textbf{43.41\%} \\
      \hline
      \multirow{2}{*}{QTacotron}
      & use bilinear interpolation & 2.11 \\
      & use FC & \textbf{2.47} \\
      \hline
      \multirow{2}{*}{QTransformer-TTS}
      & use bilinear interpolation & 2.24 \\
      & use FC & \textbf{2.32} \\
      \hline
      \multirow{2}{*}{QM5}
      & without output clip & 43.21\% \\
      & with output clip & \textbf{50.11\%} \\
      \hline
      \multirow{2}{*}{QTacotron}
      & without output clip & 2.41 \\
      & with output clip & \textbf{2.67} \\
      \hline
      \multirow{2}{*}{QTransformer-TTS}
      & without output clip & 2.39 \\
      & with output clip & \textbf{2.52} \\
      \bottomrule
    \end{tabular}
    \vskip -2mm
\end{table}

\textit{Implemented methods of LT.}
In QM5, QTacotron, and QTransformer-TTS, the LT is implemented by FC layer or the LT implemented by bilinear interpolation. As shown in Table~\ref{tab:ablation}, the results of using the FC layer are better than bilinear interpolation. In low-qubit VQC, using the FC layer outperforms the bilinear interpolation when implementing the LT.

\textit{With or without output clip.}
In QM5, QTacotron, and QTransformer-TTS, the results of LT with output clip operation are compared with LT without output clip operation. As shown in Table~\ref{tab:ablation} shows, the results of using output clip are better than without output clip. In low-qubit VQC, the results of LT with output clip operation outperform LT without output clip operation.

\section{Conclusion}
\vskip -1mm
\label{sec:conclusion}
This study introduced a novel VQC, called low-qubit VQC. It can address the Barren Plateaus challenge and lower the requirement of qubits compared to VQC. It is crucial because VQCs are used in low-qubit quantum devices in the NISQ era. Low-qubit VQC can be used to implement QNN models in low-qubit quantum devices. Based on low-qubit VQC, we propose the QSpeech toolkit. QSpeech fully uses PyTorch and PennyLane as a QML engine and significantly simplifies the implementation and training of the entire speech application. Our QSpeech provided various quantum neural layers and hybrid quantum-classical neural networks. The experimental results showed that the hybrid quantum-classical neural networks with low-qubit VQC can be successfully used in speech applications.

\section*{Acknowledgment}
\vskip -1mm
This paper is supported by the Key Research and Development Program of Guangdong Province under grant No.2021B0101400003. Corresponding author is Jianzong Wang from Ping An Technology (Shenzhen) Co., Ltd (jzwang@188.com).

\bibliographystyle{IEEEtran}
\bibliography{qspeech}

\end{document}